\documentclass[preprint,showpacs,prl,preprintnumbers,amsmath,amssymb]{revtex4-1}

\usepackage{graphicx}
\usepackage{dcolumn}
\usepackage{bm}
\usepackage{appendix}
\usepackage{float}

\begin{document}
\title{Quantum critical behavior in Ce(Fe$_{0.76}$Ru$_{0.24}$)$_2$Ge$_2$: the full story}
\author{Wouter Montfrooij$^{1,2}$ and Tom Heitmann$^{2}$}
\affiliation{$^1$Department of Physics and Astronomy, University of Missouri, Columbia, MO 65211, United States\\
$^2$Missouri Research Reactor, University of Missouri, Columbia, MO 65211, United States}%
\author{Yiming Qiu$^3$, Shannon Watson$^{3}$, Ross Erwin$^{3}$, Wangchun Chen$^{3}$, and Yang Zhao$^{3,4}$}
\affiliation{$^3$National Institute of Standards and Technology, Gaithersburg, MD 20899, United States\\
$^4$Department of Materials Science and Engineering, University of Maryland, College Park, MD, United States}
\author{Meigan Aronson$^{5}$}
\affiliation{$^5$Department of Materials Science and  Engineering, Texas A\&M University, College Station, TX 77843, United States }
\author{Yinkai Huang and Anne de Visser$^{6}$}
\affiliation{$^6$Van der Waals-Zeeman Institute, University of Amsterdam, Science Park 904, XH Amsterdam, the
Netherlands}
\begin{abstract}
{Systems with embedded magnetic ions that exhibit a competition between magnetic order and disorder down to absolute zero can display unusual low temperature behaviors of the resistivity, susceptibility, and specific heat. Moreover, the dynamic response of such a system can display hyperscaling behavior in which the relaxation back to equilibrium when an amount of energy E is given to the system at temperature T only depends on the ratio E/T.  Ce(Fe$_{0.755}$Ru$_{0.245}$)$_2$Ge$_2$ is a system that displays these behaviors. We show that these complex behaviors are rooted in a fragmentation of the magnetic lattice upon cooling caused by a distribution of local  Kondo screening temperatures, and that the hyperscaling behavior can be attributed to the flipping of the total magnetic moment of magnetic clusters that spontaneously form and order upon cooling. We present our arguments based on the review of two-decades worth of neutron scattering and transport data on this system, augmented with new polarized neutron scattering experiments.}
\end{abstract}
\maketitle 
\section{Introduction}
Magnetic ions embedded in a metal interact with each other through the conduction electrons. This interaction\cite{rkky4}, fleshed out by Ruderman and Kittel\cite{rkky1}, Kasuya\cite{rkky2}, and Yosida\cite{rkky3} (RKKY), will lead to a magneticically ordered ground state upon cooling provided the magnetic moments are not shielded by the conduction electrons. The latter process is referred to as Kondo screening\cite{kondo,kondo1}. When this screening takes precedence over the ordering tendencies, then the ground state will be non-magnetic, and it is characterized by an increased conduction electron mass\cite{sachdev}. Such a ground state is referred to as a heavy fermion ground state. The problem of isolated magnetic ions screened by the conduction electrons has been solved\cite{kondo}, however, the situation where each unit cell houses one or more magnetic ions, the so-called Kondo lattice problem\cite{hewson}, is not fully understood yet.\cite{stewart}

When the competition between the magnetic moments lining up and these selfsame moments being shielded is not resolved upon cooling the system down to absolute zero, then this system displays quantum critical behavior.\cite{sachdev} The system is on the cusp of ordering since any change in the coupling strength between the magnetic moments and the conduction electrons would see the system end up in either an ordered, or in a magnetically shielded ground state\cite{doniach}. Therefore, critical fluctuations associated with the nearby ordered phase will be present in the system, however, the decay of these fluctuations do not follow the rules of classical physics. The reason for this is that-- in finite-temperature phase transitions-- the energy cost E to create a large scale fluctuation will be less than the available thermal energy and hence, classical physics will prevail\cite{classical}. At T= 0 K this is no longer the case, and therefore, quantum physics will prevail\cite{sachdev} with the result that the approach to ordering is characterized by a set of critical exponents different\cite{heyl} from classical (finite-temperature) second-order phase transitions. Resistance, susceptibility, and specific heat measurements on metallic quantum critical systems display a response different\cite{stewart} from the predictions of Fermi liquid theory. Collectively, such unusual behavior is referred to as non-Fermi liquid behavior (nFl).

Some quantum critical systems display high-energy-like dynamical scaling, or hyper-scaling. In 1995 it was discovered\cite{meigan} that UCu$_4$Pd responded in un unusual way to being probed by neutrons depositing an amount of energy E into the system when it is held at temperature T: the dynamic susceptibility $\chi$(q,E) (the function measuring how susceptible a system is to creating a perturbation of energy E= $\hbar\omega$ and wavelength $\lambda$= 2$\pi$/q) only depended on the ratio E/T rather than on both E and T as separate variables. Later on, this behavior has also been observed in other systems\cite{meigan2,schroder,wouterprl,knafo,krish,kim}, and similar magnetic hyper-scaling has also been observed\cite{schroder} when the system is subject to an external magnetic field H (H/T scaling). Such dynamical scaling is expected for systems where the relevant internal degrees of freedom take much less energy to excite than the available thermal energy. In this case, the lifetime $\Gamma$ of all excitations is directly proportional to the inverse of the temperature, resulting in a dynamical response that only depends on E/T.

The challenges in developing an understanding of quantum critical systems are manyfold\cite{stewart}, both experimental and theoretical. From an experimental point of view, it is difficult to prepare a system to have just the right amount of competition between ordering and shielding so that it will end up at a quantum critical point (QCP) when cooled to 0 K. The competition can be fine-tuned by applying hydrostatic or chemical pressure\cite{doniach,sullow}, but the former is difficult since the pressures involved are too high for large, gram-sized samples, while the latter introduces chemical disorder rendering the interpretation of the results more cumbersome. The theoretical challenges have also proven to be formidable: despite research down promising avenues\cite{hertz,millis,neto,si,pines}, theoretical predictions either only cover a few of the observed quantum critical aspects, or they only apply to a subset of systems. Moreover, the quantum critical system that is the main topic of this paper,\cite{fontes} Ce(Fe$_{0.755}$Ru$_{0.245}$)$_2$Ge$_2$, is not captured\cite{wouterprl} by any of the leading theoretical scenarios. In addition, theoretical directions are difficult to discern when the critical exponents\cite{heyl} that describe the response of a quantum critical system close to ordering differ from system to system\cite{schroder,meigan2,knafo,kim}, rendering the search for novel universal behavior experimentally undirected.

In this paper we discuss experiments performed\cite{wouterprl,wouterprba,wouterprb,gaddy,nist} on quantum critical  Ce(Fe$_{0.755}$Ru$_{0.245}$)$_2$Ge$_2$ and we show that all unusual aspects of the response of this system can be attributed to the formation of isolated magnetic clusters upon cooling. Thus, we argue that Ce(Fe$_{0.755}$Ru$_{0.245}$)$_2$Ge$_2$ is a quantum critical system that we can fully understand based on percolation physics. We argue our point by reviewing past experiments\cite{wouterprl,wouterprba,wouterprb,gaddy} and by presenting new\cite{nist} polarized and unpolarized neutron scattering experiments that identified the low-energy excitations responsible for the observed E/T scaling behavior. In particular, we argue that the following scenario (sketched in Fig. \ref{fig1})  accounts for the observed quantum critical phenomena in  Ce(Fe$_{0.755}$Ru$_{0.245}$)$_2$Ge$_2$. 

\begin{itemize}  
\item Upon cooling, magnetic moments become increasingly more shielded by the conduction electrons; 
\item chemical doping introduces a distribution of shielding temperatures;
\item  some magnetic moments are shielded before others upon cooling and the magnetic lattice develops holes;
\item groups of magnetic moments become isolated from the rest of the magnetic lattice upon cooling and form clusters;
\item the magnetic moments in isolated clusters align with their direct neighbors because of quantum mechanical finite-size effects, while the lattice spanning cluster remains disordered;
\item  the ordering of the isolated magnetic clusters is reflected in the entropy and specific heat of the system;
\item the total spin of the isolated clusters reacts to an external magnetic field, influencing the uniform susceptibility;
\item the total spin of the isolated clusters can reorient without a cost in energy, providing the low-energy degrees of freedom necessary for dynamical scaling;
\item once the lattice spanning cluster breaks up, the system orders;
\item if the lattice spanning cluster does not break up, then the system lacks long-range order;
\item the quantum critical point is associated with the break up of the lattice spanning cluster at T= 0 K.  
\end{itemize}

The outline of the paper is as follows. We first review published experiments ('Prior Experiments' section) on  Ce(Fe$_{0.755}$Ru$_{0.245}$)$_2$Ge$_2$ and place them in the context of experiments on other systems, and in the context of the scenario outlined above. We then discuss our new experiments ('Results' section), again in the context of the scenario above. After that, we discuss the relevance of our results to other quantum critical systems, both doped and undoped ('Discussion' section).

\section{Prior Experiments on the Quantum Critical Compound}
\subsection{On Tuning the System to the Quantum Critical Point Using Chemical Pressure}
In order to ensure a compound is at the QCP, the conduction-electron mediated intermoment interaction\cite{rkky4} needs to be fine-tuned.\cite{doniach} A strong interaction leads, upon cooling, to the moments being shielded before they can line up, whereas a weak interaction leads to a magnetically ordered groundstate, provided the temperature is low compared to the effective exchange interaction between moments. Since the interaction between the moments and the conduction electrons depends exponentially on the interatomic distances,\cite{hewson,leiden} applying hydrostatic pressure or chemical pressure through the substitution of smaller or larger ions into the lattice allows for fine-tuning of the interaction strength so that at T= 0 K the system is neither ordered nor fully shielded.

While chemical doping introduces an additional level of complexity to the interpretation of experimental findings, in practice such doping does not typically lead to any substantially different critical behavior, and in addition, studying doped systems is largely unavoidable when doing scattering experiments since the required sample sizes are too large to be able to apply sufficient pressure. Also, lattice expansion can only be achieved through chemical doping. S\"{u}llow {\it et al.}\cite{sullow} showed that pushing CeRu$_2$Ge$_2$ to the QCP through pressure and through chemical doping lead to resistivity and specific heat results that were entirely consistent with each other. Furthermore, dynamical scaling has been observed in heavily doped samples,\cite{meigan, wouterprl,krish} as well as in (very) modestly\cite{schroder,knafo} doped ones. However, not all doping results in the same amount of additional complexity. For instance, replacing magnetic ions with non-magnetic ones fundamentally changes the system, whereas replacing non-magnetic ions with other non-magnetic ions only changes the interatomic distances and the number of conduction electrons. The least intrusive doping consists of replacing non-magnetic ions with iso-valent non-magnetic ions as far away as possible from the magnetic sites.  Ce(Fe$_{1-x}$Ru$_{x}$)$_2$Ge$_2$ is such a system.

CeFe$_2$Ge$_2$ is a heavy fermion system\cite{cefege} that can be driven\cite{fontes} to the QCP by expanding the tetragonal body-centered lattice through means of substituting one in four iron atoms by ruthenium. Increased substitution leads to a magnetically ordered ground state. We show the phase diagram in Fig. \ref{fig2}. During the past two decades resistivity, susceptibility, specific heat, and neutron scattering experiments have been performed on this system; we summarize the most relevant neutron scattering experiments in Table \ref{table1}.

\begin{table}
  \caption{Neutron scattering experiments on Ce(Fe$_{0.76}$Ru$_{0.24}$)$_2$Ge$_2$; the bottom two entries are new experiments for which a partial account has been given in reference \cite{nist}.}
 \label{equivalent}

  \begin{tabular}{l l c}
    \hline
   &   \\[-8pt]
  Instrument & description & reference(s) \\[3pt]
      \hline
 IN6, ILL & Established E/T scaling in polycrystal  & \cite{wouterprl,wouterprba} \\[3pt]
 HB3, ORNL & Established ordering wavevector and cluster presence & \cite{wouterprb} \\[3pt]
 HB3, ORNL & Magnetic inelastic scattering  \\[3pt]
 DCS, NIST & Established ordering wavevector and cluster presence & \cite{wouterprb} \\[3pt]
 TRIAX, MURR & Established ordering wavevector and cluster presence & \cite{wouterprb} \\[3pt]
 BT7, NIST & Established reoriention of cluster superspins & \cite{nist} \\[3pt]
 TRIAX, MURR & Established reoriention of cluster superspins &  \\[3pt]
      \hline
\end{tabular} \label{table1}
\end{table}

The resistivity, susceptibility, and specific heat measurement on single crystal  Ce(Fe$_{0.76}$Ru$_{0.24}$)$_2$Ge$_2$  show non-Fermi liquid behavior, that is, the functional behavior at low temperatures differs from the predictions of Fermi liquid theory for metals. We summarize the characteristic measurements in Fig. \ref{fig3}. These measurements show that our single crystal Ce(Fe$_{0.76}$Ru$_{0.24}$)$_2$Ge$_2$ sample is very close to the QCP (Ce(Fe$_{0.755}$Ru$_{0.245}$)$_2$Ge$_2$), albeit just on the paramagnetic side of it judging by the saturation of the data below 0.5 K.

In 2000, neutron scattering experiments on polycrystalline samples carried out on the IN6 spectrometer at Intitute Laue-Langevin (ILL) revealed\cite{wouterprl} that this system displayed E/T scaling, indicative of the presence of low-energy excitations. These experiments also revealed that the two leading scenarios of what might drive the response of a quantum critical system failed to capture the observed dynamics in Ce(Fe$_{0.76}$Ru$_{0.24}$)$_2$Ge$_2$: neither did the local susceptibility diverge\cite{si}, nor was the relaxation back to equilibrium characterized by simple exponential decay.\cite{hertz,millis} We show characteristic data in Fig. \ref{ill}.

\subsection{Formation of Magnetic Clusters in Quantum Critical Composition}

Elastic neutron scattering experiments on the HB3 triple-axis spectrometer at Oak Ridge National Laboratory (ORNL) demonstrated that  Ce(Fe$_{0.76}$Ru$_{0.24}$)$_2$Ge$_2$ is on the verge of magnetic ordering caused by a spin density wave (SDW) instability with ordering wavevector (0,0,0.45). The moments are ferromagnetically aligned perpendicular to the c-axis. As already inferred from the uniform susceptibility of this system, the moment carrying cerium ions have the tetragonal c-axis as their easy axis.\cite{loidl} The iron and ruthenium ions do not carry a local moment, so from a magnetic point of view, we have a body-centered magnetic tetragonal lattice with the {\it c}-axis being about 2.5 times as long\cite{fontes,loidl} as the {\it a}- and {\it b}-axes.

The main finding of these HB3 experiments was that, upon cooling, the magnetic lattice gets diluted and fragmented into a collection of isolated magnetic clusters.\cite{wouterprb} Thus, the system turns into a percolative system.\cite{stauffer} When the temperature gets lowered, Kondo shielding becomes more effective and cerium moments become shielded by the conduction electrons, thereby dropping out of the magnetic lattice. However, not every cerium ion gets shielded at the same temperature, most likely because of a distribution of shielding temperatures introduced by the random substitution of the larger ruthenium ions for the smaller iron ions. In this picture, the resulting distribution of interatomic separations resulted in a distribution of strengths of the interaction between the localized f-electron magnetic moment on the cerium ions and the conduction electrons, resulting in a distribution of Kondo shielding temperatures. Note that in this picture the shielding of the magnetic ions does not have to be complete: their moments merely have to become shielded enough that neighboring moments have a much weakened tendency to line up at a given temperature.

We repeat the evidence here that led to the identification of isolated magnetic clusters whose presence lies at the heart of understanding the low-temperature dynamics and transport properties of this system. Upon cooling, the magnetic scattering increases around the ordering wavevector without displaying long-range order as would be observed for resolution limited magnetic Bragg peaks. Thus, short-range order develops\cite{wouterprb} as is to be expected upon approaching a phase transition to a magnetically ordered phase. This is shown in Fig. \ref{sro}.

\subsection{Internal Ordering of Magnetic Clusters}
\subsubsection{Distribution of Kondo Temperatures}
However, the observed short-range order does not reflect critical fluctuations out of the ordered phase, rather it reveals the presence of isolated, magnetically ordered clusters. Given that the system is tetragonal, the conduction mediated interactions that drive the system to ordering have differing strengths\cite{rkky1,rkky2,rkky3,sachdev} in the ab-plane than along the c-direction. This differing strength is responsible for the moments aligning ferromagnetically perpendicular to the c-axis. As such, when evaluating the correlation lengths of the short-range order along different crystallographic directions, this direction dependent interaction strength should be reflected into direction dependent correlation lengths of the magnetic moments. Only in cases where the temperature is low enough that all available moments have lined up do we not see a dependence on interaction strength. In Ce(Fe$_{0.76}$Ru$_{0.24}$)$_2$Ge$_2$, it was observed\cite{wouterprb} that the number of correlated magnetic moments was identical along all crystallographic directions, and moreover, this was observed to be the case at all temperatures where short-range order could be observed. Therefore, the observed short-range order is not associated with (standard) critical fluctuations. Instead, we argue in the following paragraphs that  short-range order must reflect fully ordered magnetic entities that are finite in size (clusters) and that encompass equal numbers of Ce-ions along all crystallographic directions.

There are two possible explanations for the emergence of magnetic clusters in our system: it can be the consequence of a dilution of the magnetic lattice resulting from Kondo screening and turning the magnetic lattice into a percolation system, or it can be the result of stacking faults in the magnetic lattice limiting the spatial extent of critical fluctuations. In this paragraph we rule out the latter. First, on cooling we observe the correlation lengths to increase\cite{wouterprb}, but these correlation lengths span identical numbers of moments along all crystallographic directions. This holds for all temperatures probed. Stacking faults are not expected to exhibit such a temperature dependence. Second, when the Ru-concentration is slightly increased, a transition to long-range order is observed. We show this in Fig. \ref{lro}. If stacking faults were to limit the size of the correlated volume in the case of critical fluctuations, then long-range order should not emerge by changing the Ru-concentration from 0.25 to 0.26 since the magnetic lattice does not change in any essential manner. Third, if stacking faults were to be the explanation behind the observation of equal numbers of moments lining up along all crystallographic directions, then we are left with the more puzzling observation that these stacking faults are spaced in such a way as to exactly negate the c/a ratio of 2.5. In view of these three reasons-- each of which is itself sufficient to rule out attributing the peculiarities of the observed correlation lengths to stacking faults-- for the remainder of this paper we solely concentrate on the consequences of magnetic dilution caused by the aforementioned distribution of Kondo shielding temperatures.

\subsubsection{Finite Size Effects}

The reason that the moments on isolated clusters align with their neighbors is because of quantum mechanical finite-size effects. Once a sufficient number of moments have been shielded so that isolated entities (clusters) emerge, then the moments on these clusters will line up with their neighbors. The reason for this is that the wavelength of fluctuations that can disorder the alignment of the magnetic moments are restricted by quantum mechanics to fit the spatial extent of the cluster. When this is the case, then even the disordering fluctuations that require the least amount of energy come with an energy cost that is too substantial-- compared to the available thermal energy-- to arise spontaneously. This is shown in Fig. \ref{finite}. As a result, such a disordering fluctuation will not form and the moments on the cluster line up with each other.

It appears, once clusters become isolated and the moments on the clusters line up, that these clusters become protected from Kondo shielding. Experimentally this is observed\cite{wouterprba} as additional magnetic intensity around the ordering wavevector that appears, upon cooling, on top of the magnetic intensity already present at higher temperatures. That is, additional intensity is obtained at lower q, rather than spectral weight shifting from high to low q. This is shown in Fig. \ref{additional}. Thus, when new clusters peel off from the lattice spanning cluster, these newly minted clusters order and contribute to the neutron scattering signal, but the signal associated with the pre-exisitng clusters remains unchanged. Had these pre-existing clusters not been protected from Kondo shielding, then we would have expected to see their characteristic broad (in momentum space) scattering signal disappear upon cooling. Neither polycrystalline\cite{wouterprl,wouterprba} nor single-crystal experiments\cite{wouterprb} displayed this kind of behavior. From a theoretical point of view, it stands to reason that when there exists a competition between ordering and shielding, then within the ordered environment shielding becomes more difficult: Kondo shielding involves a spin-flip interaction\cite{hewson} between the electron spin and magnetic moment; when the moments cannot flip inside their ordered environment, then the shielding mechanism will be severely hampered. We will return to this putative protection from Kondo sheilding in our discussion.

\subsection{Consequences of the Formation of Ordered Magnetic Clusters}

The consequences of the emergence of isolated, magnetically ordered clusters on the transport properties and dynamical response are manyfold. The specific heat will reflect the loss of entropy of (N-1) k$_B$ ln2 when a cluster of N members in this Ising system peels off from the lattice spanning cluster. The loss is proportional to N-1 since the cluster retains its superspin\cite{vojta} degree of freedom. The closer the percolation threshold is approached, the larger the size of the clusters that peel off, and the larger the effect on the specific heat. Thus, the approach to the percolation threshold, and presumably the approach to the QCP, is marked by an increasingly rapid loss in magnetic entropy, resulting in a critical behavior of the specific heat dictated by percolation exponents.\cite{vojta,sahimi,orbach}

Isolated clusters reveal their presence both in the uniform and in the q-dependent susceptibility. Fully ordered, isolated clusters are likely to have dangling, non-compensated moments even in an antiferromagnetically ordered cluster, resulting in a non-zero superspin that can align itself with an external magnetic field. The details of the response depend on the number of uncompensated magnetic moments, which in turn depends on the number of magnetic moments per cluster as well as on the number of clusters present at a given temperature. The q-dependent susceptibility peaks at the ordering wavevector $\vec{Q}$ where all the magnetic moments contribute to the signal. The strength of this signal will be proportional to the sum over the signals from all individual clusters, whereas the signal of any individual cluster is proportional to the number s of magnetic moments squared ($\sim s^2$) that constitute that particular cluster.

The magnetic correlation length as measured in scattering experiments will be a weighted average over the sizes of all isolated clusters. This avenue has already been  pursued in an earlier publication\cite{gaddy} where Monte Carlo computer simulations were performed on a percolating system of body-centered magnetic moments. In there, the loss of entropy was evaluated as a function of site occupancy p; the specific heat c(p) corresponding to this entropy loss was then equated to the measured specific heat c(T) of  Ce(Fe$_{0.76}$Ru$_{0.24}$)$_2$Ge$_2$ in order to determine the site occupancy p as a function of temperature p(T). Then, using the latter connection between site occupancy and system temperature, the simulated magnetic correlation length $\xi$(p) could be converted into a temperature-dependent correlation length $\xi$(T). When this simulated correlation length was compared to the measured\cite{wouterprb} correlation length characterizing short-range order, a promising agreement was found. This is strong indication that the quantum critical scenario where cluster formation dictates a large part of the observed response is an avenue worth pursuing. After all, when the dynamic correlation length can be predicted, without adjustable parameters, from a direct mapping of the specific heat, then this is evidence for the fact that cluster formation plays an important and perhaps leading role in determining the approach to a QCP. We return to this issue in our 'Discussion' section.

The connection between cluster formation and the resistivity is much less clear. When isolated clusters order, then this is bound to have a repercussion on the ease with which an electron can traverse the cluster; when the moments on the cluster are ordered, then the scattering of the electron by the moments will be altered, with a reduction in resistance over the cluster the presumed outcome. The overall resistivity picture is complicated with electrons 'seeing' a lattice consisting of ordered clusters, shielded magnetic moments, as well as a backbone of the lattice spanning cluster consisting of disordered, magnetic moments. We already observed that the coherence temperature of the resistivity and the emergence of clusters as seen in neutron scattering experiments occur around the same temperature of roughly 15 K. We will show in the 'Results' section that the temperature dependences of the resitivity and cluster formation follow each other quite closely.

Inelastic neutron scattering experiments have shown that the emergence of short-range order is accompanied by a narrowing\cite{wouterprl} of the accompanying quasielastic scattering, implying that spontaneous fluctuations take increasingly longer to decay upon cooling. The highest resolution scattering experiments\cite{wouterprb} performed on DCS revealed two components. One showed that part of the scattering that develops at the lowest temperature becomes resolution limited, implying that there exist fluctuations below 0.5 K that take at least 80 ps to relax back to equilibrium. The other showed quasielastic scattering whose energy width was much larger, implying that the characteristic relaxation time is of the order of 1 ps. In all these prior experiments, the magnetic scattering was identified through a straightforward subtraction of the signal at high temperatures (T $\sim$ 50 K) from the low temperature signal. However, this procedure is not accurate enough, owing to the incoherent scattering of the sample and the strong temperature dependence of the phonon contribution, to unambiguously identify the character of the magnetic scattering.

In summary, the prior experiments have established the following for quantum critical  Ce(Fe$_{0.76}$Ru$_{0.24}$)$_2$Ge$_2$. The system displays quantum critical behavior in resistivity, specific, heat, and suscpetibility measurements. Neutron scattering experiments demonstrate that short-range order emerges around the coherence temperature of the resisitvity curve, and that the magnetic correlations in this easy-axis body-centered system of Ce-ions are attributable to the emergence of magnetic clusters that form upon dilution of the magnetic lattice. The system exhibits E/T-scaling that cannot be explained by a spin density wave instability\cite{hertz,millis} or in terms of the local moment scenario.\cite{si} In this paper, we argue that the E/T-scaling is attributable to a new degree of freedom that is inherent to a magnetic cluster of fully aligned magnetic moments. The total magnetic moment of such a cluster, called a superspin,\cite{vojta} can point up or down along the easy axis. A reorientation of this spin (superspin flipping) should not cost any energy, thereby providing the system with the low-energy degree of freedom that is necessary to endow the system with the requirements for E/T-scaling. The polarized neutron scattering experiments we discuss in the 'Results' section demonstrate that superspin flipping indeed takes place.

\section{Results}
Polarized neutron scattering experiments separate the signal attributable to magnetic excitations from the signal due to non-magnetic scattering. We set up our BT7 polarized neutron scattering experiments in such a way that all magnetic scattering would show up in the spin-flip channel (see Appendix). The spin-flip channel measures neutrons that were scattered while undergoing a flip of their intrinsic angular momentum, a process that only occurs during magnetic scattering. Using this set up, we were able to scrutinize the three components to the magnetic scattering: the scattering associated with the incipient clusters, the scattering associated with moments that are not located on clusters and that shows up as a broad background, and the magnetic scattering associated with the interaction between the neutron and the nuclei of the atoms (the so-called incoherent scattering\cite{scattering}).

We first look at the two magnetic components that are not associated with cluster formation. The earlier experiments could not distinguish whether the broad background that shows up at all momentum transfers was due to nuclear or magnetic processes. The BT7 experiments discussed here show that the broad background of quasielastic scattering persists down to the lowest temperatures. We demonstrate this in Fig. \ref{allmagnetic} where we show the spin-flip channel, which is only sensitive to magnetic scattering. This component of the scattering is seen to obey the detailed balance principle as demonstrated by the scattered intensity at negative energy transfers. Given this persistent background signal, this finding implies that neither all of the magnetic moments become shielded nor end up in fully ordered isolated clusters.  The third component is the magnetic incoherent scattering whose intensity is distributed over the spin-flip (2/3) and non-flip (1/3) channels\cite{polscat}. This scattering is temperature independent and is isolated by comparing scattering data at the temperature of interest to higher temperatures; as such, this (unwanted) scattering does not stand in the way of the interpretation of the data.

We show the temperature dependence of the quasielastic scattering at the ordering wavevector Q= (0,0,0.45) in Fig. \ref{quasi}. The data in this figure were obtained in the spin-flip channel, and as such, these data orginate from magnetic scattering by the sample. Earlier neutron scattering experiments\cite{wouterprl,wouterprb} have shown that short-range order develops upon cooling down Ce(Fe$_{0.76}$Ru$_{0.24}$)$_2$Ge$_2$: scattered intensity starts to pile up around the ordering wavevector on top of a broad background of scattering. The magnetic signal of our polarized experiments shows a temperature dependence consistent with these earlier, non-polarized experiments and with the interpretation of isolated clusters emerging and surviving upon cooling. As can be seen in Fig. \ref{quasi}, the scattering increases with decreasing temperature for all energy transfers E$>$ 0. As discussed in the previous section, this behavior is what is expected for increasingly more and increasingly larger clusters becoming isolated upon cooling, with the clusters being protected from further Kondo shielding because of their intrinsic ordering. Thus, polycrystalline experiments, single crystal non-polarized and single crystal polarized experiments all clearly demonstrate that the magnetic scattering associated with short-range order exhibits a behavior that is different from standard critical scattering upon approaching a phase transition.

Our time-of-flight data obtained on the high-resolution DCS spectrometer support the scenario laid down in the previous paragraph. Time-of-flight neutron scattering experiments measure scattering in hundreds of detectors simultaneously: a pulse of neutrons hits the sample, and the neutrons are detected over many scattering angles as a function of their time of flight. This flight time is readily converted to the amount of energy the neutron has imparted onto the sample. Since the detectors are fixed in space, the data are not acquired at constant $\vec{q}$; rather, cuts through ($\vec{q}$,E) space have to be made to access the subset of data that are of most interest. We show such subsets in Figs. \ref{dcsord} and \ref{dcsenergy} for scattering around the ordering wavevector $\vec{Q}$. The details of the data reduction procedure are in the captions and in the Appendix. The cuts at constant energy E=0 meV (Fig. \ref{dcsord}) show that the scattering increases upon cooling, and that the increase in scattering at the lowest temperatures is such that additional scattering appears on top of existing scattering (existing at higher temperatures). Note that these data cannot be compared directly to triple-axis data since {\it h,k}, and {\it l} all vary, but it is clear that the scattering is consistent with the cluster scenario outlined earlier.

Combining the DCS experiments with the polarized experiments we can isolate the dynamics of the clusters rather than merely establishing their presence. We do this in the next few paragraphs. Cuts through the DCS data displaying their energy dependence (see Fig. \ref{dcsenergy}) near the ordering wavevector show that the additional scattering at the lowest temperatures (T $<$ 1 K)  is resolution limited. Thus, either this additional scattering is elastic in origin and associated with frozen-in superspin, or it corresponds to re-orientations of the superspins on time scales $\Delta$t exceeding 80 ps. The figure of 80 ps follows from the energy resolution ($\Delta$E= 0.05 meV= 0.076 ps$^{-1}$ for $\lambda_{inc.}= 4.8$\AA) of the DCS spectrometer: $\Delta$t= 2$\pi$/$\Delta$E.

Additional insight in the nature of the developing short-range order can be gained by comparing the temperature dependence at the ordering vector for the elastic channel and an inelastic channel. The former shows the scattering associated with (short-range) ordered moments of clusters whose superspin is frozen in (at least on the time scale of the experiment), while the latter shows the scattering associated with fluctuating moments (both individual moments and the superspins of clusters). We show the results in Fig. \ref{ordscat}. As can be seen in this figure, the elastic and inelastic magnetic scattering data display a qualitatively different temperature dependence. We argue in the 'Discussion' section that this difference is what is to be expected when clusters form whose superspins can fluctuate but where the largest clusters freeze out due to the presence of dissipation, as predicted by Hoyos and Vojta.\cite{vojta}

Next we show that the width in reciprocal space of the short-range order is smaller in the elastic channel than in the inelastic channel. We argue in the 'Discussion' section that this observation is consistent with the short-range order scattering originating from isolated clusters. The width in reciprocal space of the short-range scattering for the elastic channel has already been measured with great precision\cite{wouterprb} on HB3, DCS, and TRIAX. The width was found to be 0.166 r.l.u. (reciprocal lattice units) at T= 2 K. We have used the polarized capability of BT7 to determine the correlation length at E= 1.25 meV, and found that the inelastic correlation length is about twice as short (corresponding to roughly double the width in reciprocal space). We show this in Fig. \ref{inelcorlength}.  We attempted to measure the reciprocal width at higher energy transfers as well, but because of time restrictions on BT7 and nuclear scattering contamination in non-polarized experiments on TRIAX we were unable to discern the weaker signal at 2.5 meV (polarized) and at various energy transfers (unpolarized) from the background, at least not with sufficient accuracy that we could determine its width. We show our attempts using unpolarized scattering for a number of energy transfers in Fig. \ref{failed}. The attempts employing unpolarized scattering for T $<$ 5 K, using the higher temperature background subtraction method, failed because the temperature dependence of the phonon contribution was too strong over the region of interest.

The onset of short-range order occurs roughly at the coherence temperature of the resistivity, at about 15 K. The coherence temperature is viewed\cite{sachdev,hewson} as the temperature where Kondo shielding takes place resulting in a sharp reduction of the resistance. Since the shielding mechanism is responsible for the emergence of magnetic clusters, it stands to reason that the onset of short-range order and the drop in resistivity occur around the same temperature. However, our results point toward a slightly different interpretation of exactly what causes the drop in resistivity. Given that we only observe short-range order once clusters become isolated, the similarity between the coherence temperature and the emergence of magnetic clusters implies that the coherence temperature should not be associated with the onset of Kondo shielding, but rather with the emergence of magnetic clusters. The latter only occurs well into the Kondo shielding process, not at its onset. Of course, in a system without a distribution of shielding temperatures this would not be a point of discussion. We detail the connection between the resistivity and the neutron scattering data in the next paragraph.

The scattering intensity at the ordering wavevector and the inverse of the resistivity display a strong correlation, as we show in Fig. \ref{resis}. We took the measured scattered intensity at the equivalent ordering wavevector (1,1,0.45) as obtained on TRIAX and HB3 (see Fig. \ref{ordscat}), and normalized them relative to each other to coincide for T $>$ 10 K to account for the larger incident neutron flux on HB3 and the slightly different monitor and detector efficiencies. The two spectrometers were operated at an almost identical energy resolution so that we can directly compare the temperature dependence of the data as measured on the two spectrometers. Fig. \ref{resis} shows that the two data sets agree well with each other. For our resistivity data, we subtract the measured resistivity at the lowest temperatures ($\rho_0$), and we plot 1/($\rho$(T)-$\rho_0$) in the same graph. We multiply 1/($\rho$(T)-$\rho_0$)  by a scale factor after we subtracted the value of this quantity at 100 K; the scale factor is there to account for the fact that electrical resistivity and neutron scattering counts do not have much in common. We emphasize here that we do not attempt any quantitative agreement but rather demonstrate a correlation in the temperature dependence of these two independent characterizations of the system response. Bearing that in mind, Fig. \ref{resis} indicates that the marked temperature dependence of the resistivity around the coherence temperature and the onset and temperature dependence of short-range order are indeed closely correlated. As such, our data are consistent with the scenario where the coherence temperature is not related to Kondo screening by itself, but to the emergence of fully ordered, isolated clusters. Presumably these isolated clusters present a path of much lower resistance caused by the ordered nature of the magnetic moments in the cluster.

\section{Discussion}
In this discussion we make our case for the quantum critical point in Ce(Fe$_{0.76}$Ru$_{0.24}$)$_2$Ge$_2$ being the point where the lattice spanning cluster first survives down to zero Kelvin, and that the unusual non-Fermi liquid behavior, including E/T-scaling, can be ascribed to isolated, yet fully ordered magnetic clusters. After that we discuss how this scenario might be relevant to other doped systems, and discuss whether stoichiometric systems might also be prone to harboring magnetic clusters.
\subsection{Cluster dynamics}

The fact that Kondo shielding takes place in Ce(Fe$_{0.76}$Ru$_{0.24}$)$_2$Ge$_2$ was already known from the observed diminishing of the local moments with decreasing temperature\cite{wouterprl} as observed in the uniform susceptibility and the scattering by the polycrystalline sample,\cite{wouterprl} as well as in the small size of the ordered moment (0.18$\pm$ 0.03$\mu_B$/Ce ion\cite{wouterprb}). It is also known from the study of other systems that chemical substitution leads to a distribution of Kondo temperatures. As such, we can take Kondo shielding in the presence of a Kondo distribution as an established fact in Ce(Fe$_{0.76}$Ru$_{0.24}$)$_2$Ge$_2$. We sketch in Fig. \ref{sub} why substituting Fe ions for larger Ru ions locally leads to a suppression of the shielding tendencies. When a system harbors a distribution of shielding temperatures, then this must necessarily lead to a diluted magnetic lattice upon cooling. Note that Kondo shielding of a particular moment does not have to be complete: shielding merely has to weaken the local moments sufficiently so that the interaction with the neighbors becomes weaker than the available thermal energy. When this happens and clusters form, then the moments have to align according to quantum mechanics.

What cannot be reasoned out using basic physics is whether the moments on isolated clusters are protected from further Kondo shielding, or not. There are multiple aspects to consider. First, while we know that Kondo shielding involves a spin-flip interaction\cite{kondo} which probably will be more difficult to accomplish within the ordered environment of a cluster, we also know that this shielding is a collective effect\cite{hewson} involving a cloud of electrons, and as such, oversimplified imagery should not be used to justify the interpretation of clusters being protected from further shielding. Second, even if the ordered environment were to protect the moments from further demise, it would only be a metastable configuration if the true lowest energy state is that of all moments being shielded. As such, it should  be possible to tunnel into this lower energy state. Third, it is possible that the 1 in 4 substitution of Ru for Fe ions has created 1 in 4 Ce moments whose Kondo temperatures have been lowered to such an extent that they will never be shielded at experimentally accessible temperatures. Should this be the case, then it is clear that a large fraction of the moments on the clusters will never be shielded, independent of whether the cluster is ordered, or not.

Given these considerations, we only rely on experimental observations as to whether clusters dissipate on cooling, or not. For our interpretation of the data, we need to look at the temperature dependence of the dynamic susceptibility rather than that of the dynamic structure factor.\cite{scattering} The dynamic structure factor S(q,E) incorporates the Bose population factor and satisfies the detailed balance condition. Since the energy integral over S(q,E) yields the static structure factor, which only varies slowly with temperature, lowering the temperature of the system sees scattering intensity pushed from negative energy transfers to positive energy transfers. This transfer of intensity makes the interpretation more complicated, especially for a disordered system. When we look at the dynamic susceptibility, however, then the effects of the Bose population factor have been taken out of the analysis and a clearer picture emerges. We do this in the following.

The imaginary part $\chi"(q.E)$ of the dynamic susceptibility $\chi$ follows from the dynamic structure factor as\cite{scattering} 
\begin{equation}
\chi"(q,E)=(1-e^{-E/k_BT})S(q,E)
\end{equation}
with $k_B$ Boltzmann's constant. $\chi"$ is an odd function in energy, and as such, is easier to interpret the symmetric function $\chi"(q,E)/E$ since both negative and positive energy transfers can then be readily compared. This function is directly available from neutron scattering experiments provided they have been fully corrected for background effects. When using triple-axis spectrometers this correction is more difficult to perform than when using time-of-flight measurements where the detectors are fixed in space and better shielded. As such, we focus on the data on polycrystalline Ce(Fe$_{0.76}$Ru$_{0.24}$)$_2$Ge$_2$ obtained using the IN6 spectrometer at the ILL.

 We first revisit Fig. \ref{ill} in order to see whether all the data shown are consistent with the scenario of clusters peeling off of the infinite cluster upon cooling and these newly minted clusters surviving (additional) Kondo shielding upon further cooling. In this scenario, the smaller clusters are the first (upon cooling) to form. The signal of these clusters in scattering experiments is broad in momentum transfer (reflecting their small size), and broad in energy (reflecting that small clusters reorient rapidly). Thus, at the higher temperatures we expect to see a similar temperature dependence close to ordering ($|\vec{Q}|$= 3 nm$^{-1}$) and away from ordering. The insets of Fig. \ref{ill} are consistent with this anticipated behavior with a similar signal strength appearing for T$>$ 10 K for both momentum transfers. When we cool down further, larger clusters become isolated. The signal of these clusters is more narrow in momentum transfer (since they are larger) as well as in energy transfer (since it will take longer for them to reorient). Thus, we expect new scattering to appear closer to the ordering wavevector that is more narrow in energy. As such, the data away from  $|\vec{Q}|$ should start to become temperature independent for the case that the already existing clusters survive, and should diminish in intensity should the existing clusters still be subject to shielding. The data shown in the bottom half of Fig. \ref{ill} support the former. Moreover, we also observe that the data at $|\vec{Q}|$ become temperature independent at the lowest temperatures for E$>$ 0.5 meV, while the signal closer to E= 0 continues to increase upon cooling. Again, this is exactly as anticipated for larger clusters forming while smaller clusters are no longer subject to Kondo shielding.

When we take constant energy cuts through the data we observe a similar pattern: away from ordering (that is, away from q=$|\vec{Q}|$, E=0) the signal saturates upon cooling, while close to ordering the signal continues to increase. We show these findings in Fig. \ref{in6}. At an  energy transfer of E= 1.25 meV we observe that the signal has saturated for all momentum transfers, indicating that while the clusters that formed at higher temperatures can still fluctuate, the newly emerging clusters fluctuate on time scales longer than 2$\pi$/1.25 meV= 3.3 ps. We have to bear in mind, though, that because of the path that the detectors cut out in phase space, the ordering wavevector was not covered by this cut. Also note that the signal is broad in momentum transfer, reinforcing the interpretation that rapidly fluctuating clusters are limited in their spatial extent.

The data at E= 0.5 meV (top right panel Fig. \ref{in6}) are considerably more narrow in momentum transfer, indicating that this signal is dominated by larger clusters (larger than at E= 1.25 meV). Note that the cuts at E= 0.5 and E= 1.25 meV  (bottom left and right panels of Fig. \ref{in6}, respectively) do not share the same vertical scale factor; thus, what may appear to be a saturation of the signal for q $>$ 5 nm$^{-1}$ for T $<$ 7.5 K, in fact does have some minor temperature dependence to it. Also note that the energy window of the cut was fairly large in both cases, and therefore, we should compare q values within a single cut rather than compare the q-values between the cuts at different energies. At E= 0.5 meV, we observe that the signal saturates at the lowest temperatures, indicating that the clusters that do form at the lowest temperatures are predominantly larger, resulting in a fluctuation rate that does not add to the scattering at E= 0.5 meV (corresponding to roughly 8 ps). In contrast, we can see that the scattering at E= 0 meV continues to increase with decreasing temperature, indicating that clusters are still being formed, but these clusters are so large (as evidenced by their narrow width in momentum transfer) that they do not fluctuate on time scales shorter than 8 ps.

The data shown in Figs. \ref{ordscat} and \ref{inelcorlength} confirm the cluster dynamics. Fig. \ref{ordscat} demonstrates that clusters are forming at all temperatures as indicated by the increase in elastic scattering, but the clusters that form at the lowest temperatures fluctuate so slowly (or not at all), that they do not contribute to the scattering at E= 1.25 meV. Thus, the anticipated cluster distribution close to the percolation threshold provides a natural explanation for the diverging behavior observed in Fig. \ref{ordscat}. Fig. \ref{inelcorlength} affirms this picture: the scattering at E= 1.25 meV is caused by the smaller clusters, and as such, it is relatively broad in momentum transfer. The scattering at E= 0 meV arises from all clusters with the strongest contributions coming from the largest clusters, and as such, this scattering is more narrow in momentum transfer.

Overall, we observe that clusters are present at all temperatures, and that the smaller ones of these clusters are not frozen in. As such, cluster reorientations provide a low energy degree of freedom for the system. We identify this degree of freedom as the one that is responsible for bestowing E/T-scaling on to the system. This is based on the observed presence\cite{wouterprl} of E/T-scaling, something which can only happen if low energy degrees of freedom are present and on the fact that our neutron scattering investigation has not found any other low energy degrees of freedom. We note that even in a classical system E/T-scaling has been observed\cite{heitmannprb} as a result of the presence of magnetic clusters. Lastly, we find no evidence for the moments within clusters being shielded upon further cooling, and therefore, we conclude that isolated clusters are protected from Kondo shielding. Whether this is because of the moments being immune from Kondo shielding, because of the Ru substitution, or whether it is because the ordered environment precludes Kondo screening we cannot infer from our data.

\subsection{Predictions of the percolation scenario}
The scenario in which isolated clusters form, order, and are protected from further screening upon cooling comes with specific predictions for the behavior of heavily doped quantum critical systems. We discuss these predictions in this section.

\begin{figure}
\begin{center}  
\includegraphics*[viewport=60 100 450 480,width=85mm,clip]{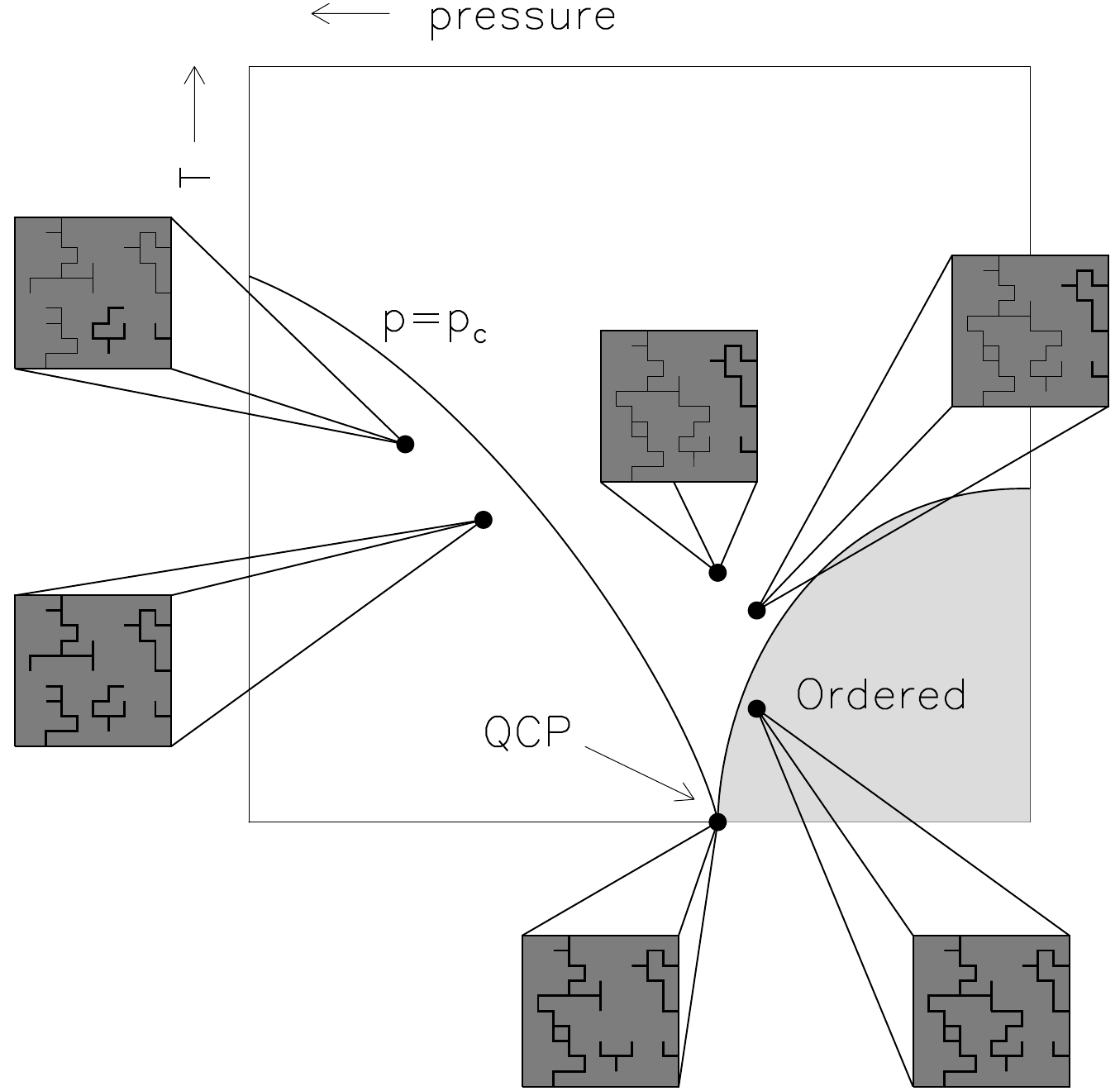}
\end{center}
\caption{Putative phase diagram for a quantum critical system subject to chemical disorder.
A magnetically ordered phase (grey area) can be reached by cooling down a classical system. Upon applying pressure, be it hydrostatic or chemical pressure, the transition temperature is driven down to 0 K, the quantum critical point. In the case of systems subject to chemical disorder, a distribution of Kondo-shielding temperatures will result in the formation of magnetic clusters upon cooling because of the temperature dependent moment surviving probability $p$. The moments within isolated clusters will line up at a temperature inversely proportional to the size of the cluster. This is shown in the insets of the figure where the square box depicts the whole lattice, and the lines depict surviving neighboring moments. Thick lines indicate that the moments within such clusters have ordered, thin lines indicate that the moments have not lined up yet because of thermal fluctuations. The lattice spanning cluster that exists above the percolation threshold ($p>p_c$) connects the top to the bottom of the lattice. Once the moments within a cluster have lined up, they are less likely to suffer from Kondo shielding. In this picture, the QCP is the point where the lattice spanning cluster survives all the way down to zero Kelvin while maintaining its capability of ordering. This infinite cluster can be broken up by removing a single moment.} 
\label{fig1}
\end{figure}

\begin{figure}
\begin{center}  
\includegraphics*[viewport=110 150 430 355,width=85mm,clip]{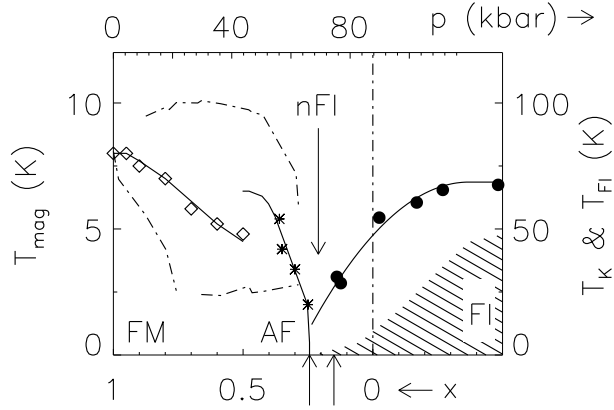}
\end{center}
\caption{Ce(Fe$_{1-x}$Ru$_{x}$)$_2$Ge$_2$ phase diagram as a function of doping (bottom axis) and pressure (top axis). This figure has been adapted from Ref. \protect\onlinecite{wouterprl}. The end compound CeRu$_2$Ge$_2$ is ferromagnetic\protect\cite{fontes} (FM) with a transition temperature of about 8 K (left axis). The other end compound CeFe$_2$Ge$_2$ is a heavy fermion system\protect\cite{cefege} (dashed-dotted vertical line). Upon replacing Ru with Fe, the FM transition temperature drops \protect\cite{fontes} (open diamonds). Once more than 50\% of the Ru has been replaced with Fe the system exhibits an antiferromagnetic (AF) phase (stars) \protect\cite{wouterprl} and reaches a QCP at x= 0.245. Non-Fermi liquid (nFl) behavior has been observed at this concentration\protect\cite{sullow}. When the   CeRu$_2$Ge$_2$ compound is subject to hydrostatic pressure (top axis), then a similar pattern is observed (curve dashed-dotted lines) \protect\cite{sullow}. Upon applying increasingly more pressure, the system is driven through the QCP and a Fermi liquid (Fl) phase is recovered (shaded area where the resistivity exhibits a $\sim$ T$^2$ behavior \protect\cite{sullow}). Fermi liquid and Kondo shielding temperatures \protect\cite{sullow} for the pressurized system are displayed on the right vertical axis. All solid lines are guides to the eye.} 
\label{fig2}
\end{figure}

In our scenario, the quantum critical point is the point in the phase diagram where the lattice spanning cluster can first survive down to zero Kelvin (see Fig. \ref{fig1}). Any finite-sized clusters that peel off must necessarily order upon cooling, and thus, true long-range order can only be achieved when the infinite cluster does not break up. Approaching the quantum critical point by moving along the phase diagram at zero Kelvin (see Fig. \ref{fig1}), we do see an increase in magnetic correlation length associated with larger and larger cluster fragments becoming isolated when the system has more and more moments that cannot be Kondo shielded (that is, with increased Ru substitution). Very close to the QCP, the size of these fragments diverges and the correlation length diverges along with it. This is sketched in Fig.  \ref{fig1}.

When clusters peel off and order, the loss of magnetic entropy $\Delta S$ (in an Ising system) is given by (s-1)k$_B$ ln2 when a cluster (endowed with a superspin) of s members peels off. The critical behavior of the infinite cluster close to the percolation threshold $p_c$  is given by\cite{stauffer}
\begin{equation}
P(p)=P_0(p-p_c)^{\beta}+(p-p_c)
\label{perc1}
\end{equation}
In here, $p$ the the occupancy (fraction of surviving magnetic moments), $P(p)$ is the fraction of the lattice sites that is part of the infinite cluster, $P_0$ is a universal\cite{stauffer} pre-factor and the exponent $\beta$ is the universal exponent pertinent to the percolation problem. In our case, $p >p_c$ since there is no critical behavior once the lattice spanning cluster has split up into fully ordered pieces. Taking the derivative of eqn. \ref{perc1} for when we remove a single site from the infinite cluster in a lattice with N sites (N$\Delta p$= -1) we find
\begin{equation}
\frac{dP(p)}{dp}=\frac{-N \Delta P(p)}{-N \Delta p}= \frac{s}{1}=\beta P_0(p-p_c)^{\beta-1}+1
\label{perc2a}
\end{equation}
Bearing in mind that isolated clusters are protected from shielding, and therefore, that every site removal has to take place on the infinite cluster, eqn. \ref{perc2a} leads to 
\begin{equation}
\Delta S \sim (s-1)  \sim (p-p_c)^{\beta-1}
\label{perc3}
\end{equation}
This equation is readily integrated to yield
\begin{equation}
S(p) \sim (p-p_c)^{\beta}
\label{perc2}
\end{equation}
Thus, close to the QCP, the critical behavior of the entropy is given by the critical exponent of the associated percolation problem. In fact, the magnetic entropy follows the strength of the infinite cluster. The critical behavior of the entropy is readily accessible from experiments, however, it is obtained as a function of temperature, not as a function of magnetic-site survival rate. The critical behavior as a function of temperature will be given by a convolution of eqn. \ref{perc2} and the Kondo shielding distribution p(T). Once p(T) is known, then the critical behavior can be compared to eqn. \ref{perc2}. As an aside, the exponent is given by $\beta$= 0.41 for a percolation scenario where the clusters are not protected,\cite{stauffer} and it is given by $\beta/(\beta+1)$ =0.27 for the case\cite{heitmannonline} where the clusters are protected.

In practice-- and as already mentioned in the introduction-- equation \ref{perc2} can be used to determine the actual distribution of Kondo shielding temperatures p(T) from specific heat experiments. P(p) can be determined accurately from simulations and from it S(p). And then, S(p) from simulations can be equated to S(T) from experiments to infer p(T). This is independent of any particular value of the critical exponent, but it does rely on a somewhat subjective separation\cite{gaddy} between the magnetic entropy associated with ordering and the entropy associated with Kondo shielding in the absence cluster formation.

Percolation theory defines additional critical exponents as follows.\cite{stauffer} Defining the cluster distribution function $n_s(p)$ as the chance that  a lattice site belongs to a cluster with s members, the  mean cluster size M(p) is given by (close to the percolation threshold):
\begin{equation}
M(p)= \sum_s s^2n_s(p)/p_c \sim |p-p_c|^{-\gamma}
\end{equation}
The divergence of the correlation length $\xi$ is given by the exponent $\nu$:
\begin{equation}
\xi(p) \sim |p-p_c|^{-\nu}
\end{equation}
In scattering experiments, the strength of the signal at E= 0 and q= $|\vec{Q}|$ corresponds to the mean cluster size M, provided both the energy and angular resolution of the spectrometer are much sharper than the intrinsic width of the signal in (q,E)-space. Neutron scattering is an interference technique, and therefore, when all magnetic moments are probed with the same phase as is done when we do elastic measurements at the ordering wavevector, then the signal of an individual cluster is proportional to $\sim s^2$, and the overall signal is weighted by the cluster distribution function $n_s(p)$. Thus, the staggered susceptibility $\chi_{\vec{Q}}$ close to the QCP is given by 
\begin{equation}
\chi_{\vec{Q}}(p) \sim |p-p_c|^{-\gamma}
\label{staggered}
\end{equation}
Both eqns.  \ref{perc2} and \ref{staggered} express the critical behavior of an experimentally accessible quantity as a function of p-p$_c$, or in our case, as a function of temperature when we substitute the temperature dependence of the shielding. Thus, the critical behavior, as a function of temperature, of the entropy and of the staggered susceptibility is given by (with $f(T)\equiv$ p(T)-p$_c$)
\begin{equation}
\chi_{\vec{Q}}(T) \sim |p(T)-p_c|^{-\gamma}= f(T)^{-\gamma}; ln \chi_{\vec{Q}}(T)=\alpha -\gamma lnf(T)
\end{equation}
\begin{equation}
S(T )\sim |p(T)-p_c|^{\beta}= f(T)^{\beta}; ln S(T)=\mu +\beta lnf(T)
\label{bobo}
\end{equation}
Thus, we can plot both $lnS(T)$ and $ln\chi_{\vec{Q}}$(T) versus temperature. The two curves should be offset from each other and their temperature variation should only differ by a factor of -$\gamma$/$\beta$ in the critical region. Provided both quantities can be determined with sufficient accuracy from experiment, this provides a way to determine both p(T) from experiment as well as the ratio of the critical exponents. We have not attempted to do so given the inaccuracies that emerge when trying to isolate the magnetic entropy from the overall entropy, and given that our computer simulations are not accurate enough yet to determine $\gamma$. We are currently working on improving the accuracy of our simulations.

The uniform susceptibility (q= 0) will also follow the cluster distribution, but in a non-ferromagnetically coupled system the connection to the cluster distribution function will be obfuscated by the size of the net moment on a particular cluster. When magnetizing a sample, it is the net moment of a cluster that will be susceptible to the external magnetic field; however, the net moment is determined by chance. The net moment depends on how many uncompensated moments there are on a cluster, which we expect to be a distribution based on the cluster size. Thus, the net moment at any given temperature as probed in a susceptibility experiment should be given by a convolution of the distribution of cluster sizes, and the distribution of dangling moments for a given cluster size. Simulations will have to be performed before we can make the connection to experimental results. In addition, the uniform susceptibility might well be dominated by those rare clusters\cite{neto,vojta} that happen to have many uncompensated moments. Such clusters do not necessarily have to be the largest clusters. As such, it is not clear whether a relationship such as eqn. \ref{bobo} exists for the uniform susceptibility.

Another interesting aspect to consider is the actual amount of doping that would be required to drive a system to the percolation threshold. The percolation threshold for a body-centered lattice is 0.246\cite{sahimi}, a value identical (within the experimental error of less than 0.01) to the observed critical concentration of Ru required\cite{fontes} to drive CeFe$_2$Ge$_2$ to the QCP. Assuming that the substitution of one Ru ion for one Fe ion would lead to one Ce moment being less likely to undergo Kondo shielding, the connection appears to be straightforward. However, we have to view this in the light of the experimental finding that isolated clusters are protected from further shielding. When this is the case, then all moments that are removed are being removed from the infinite cluster. As a result, this cluster breaks up at a higher concentration, and the relevant percolation threshold is shifted\cite{heitmannonline} to 0.272.  Therefore, these numbers do not appear to match, and could even be considered to be indicative of the percolation scenario described in this paper as being incorrect. As it turns out, we would still expect a critical concentration of 0.246 to be associated with the QCP, as we argue next.

In standard percolation, any moment can be removed and as long as 24.6 \% of the moments remain, then a percolating, infinite cluster will be present. At first sight, one would expect this critical concentration to be different in the scenario where only moments from the infinite cluster can be removed: if 24.6 \% of moments that cannot be shielded is already enough to produce a lattice spanning cluster, then should having protected moments in isolated clusters in addition to these unshieldable moments not lower this requirement? This reasoning breaks down for two reasons. First, close to the percolation threshold the surviving moments are not members of the percolating cluster, rather they reside in isolated clusters. As eqn. \ref{perc1} shows, the weight P(p) of the infinite cluster goes to zero at the threshold.\cite{stauffer} Another way of realizing this is that the percolating cluster at the threshold is a fractal,\cite{orbach} so its weight goes to zero in a three dimensional structure. Second, in the protected-cluster case, these isolated clusters have more members than in standard percolation which exactly negates the anticipated difference. Thus, while some 3\% more of the moments need to remain unshielded, this number is reached automatically by having 24.6\% unshieldable moments combine with the 3\% of shieldable moments that end up in protected clusters. An interesting aside is that both in Ce(Fe$_{0.76}$Ru$_{0.24}$)$_2$Ge$_2$ and in UCu$_4$Pd (as discussed in the next section), the impurity ions (Ru and Pd, respectively) form a fractal network at all temperatures.

\subsection{Conclusions}
We have presented a scenario in which the quantum critical behavior in Ce(Fe$_{0.76}$Ru$_{0.24}$)$_2$Ge$_2$ is ascribed to the emergence of isolated magnetic clusters that form spontaneously upon cooling. This scenario is rooted in the distribution of Kondo temperatures associated with random doping of the system, and in basic quantum mechanics that dictate that fluctuations have to match the size of their environment. We have reviewed the evidence for the presence of such clusters, and have presented new data showing that these clusters have a superspin that can account for the observed E/T hyper-scaling in the absence of local criticality. Where our data were not accurate enough to draw direct conclusions, we have shown that they are consistent with the cluster scenario. For instance, the anticipated increased width in momentum transfer with increased energy transfer was consistent with our data, as well as the temperature dependent saturation of the scattering away from the ordering wavevector.

We have discussed the connections that exist in this scenario between the susceptibility, specific heat, and the magnetic correlation length. These connections come with testible predictions, but they require accurate computer simulations as a go-between. Our initial simulations\cite{gaddy,heitmannonline} do support the existence of the proposed connections. We have shown (Fig. \ref{resis}) that there exists a connection between the resistivity and the emergence of clusters; we have not attempted to quantify this connection given the complexity of the resitivity in a system that has shielded moments, unshielded moments free to fluctuate, and unshielded moments locked up in clusters. We are currently performing computer simulations to develop a better understanding of the resistivity in the cluster scenario. In summary, the dominant contribution to the non-Fermi liquid behavior observed in Ce(Fe$_{0.76}$Ru$_{0.24}$)$_2$Ge$_2$ has been identified and all aspects of this non-Fermi liquid behavior are satisfactorily explained by the emergence of clusters, at least at the qualitative level.

We end this section with a few remarks on whether we expect the cluster scenario to be relevant to other doped systems, and perhaps even to (nearly) stoichiometric systems where non-Fermi liquid behavior has been recorded. These remarks should be viewed as speculation and as avenues for future research, avenues that we are currently pursuing. Up front, we mention that we have no evidence of this scenario playing out in other quantum critical systems. The reason for this is not the existence of data to the contrary, but the lack of data concerning measurements of magnetic correlation lengths along different crystallographic directions.

The first system in which E/T-scaling was observed\cite{meigan} is UCu$_4$Pd. The uranium moments form a face-centered structure, and susceptibility\cite{bernal2} and muon measurements\cite{bernal1,bernal3} have shown that the Pd substitution on the Cu sites is random and leads to a distribution in Kondo temperatures. As such, this heavily doped system appears to be a good candidate to compare it to the cluster scenario. However, the system has a cubic structure so unlike the case for Ce(Fe$_{0.76}$Ru$_{0.24}$)$_2$Ge$_2$, identical magnetic correlation lengths along different cyrstallographic directions cannot be interpreted as evidence in favor of our cluster scenario since the interaction strengths are identical along different directions as well. Also we are not aware of single crystal neutron scattering experiments having been performed on this system. Speculating though, if the cluster scenario were to hold for UCu$_4$Pd, then we would expect the critical doping concentration to be given by the percolation threshold for a face-centered system. This threshold is 0.198\cite{sahimi}, identical (within experimental error) to the 0.2 Pd substitution on the Cu sites. Of course, this comparison only makes sense if one Pd substitution leads to one U moment that will not be shielded.

Even when it is correct that a 1 in 4 substitution in Ce(Fe$_{0.76}$Ru$_{0.24}$)$_2$Ge$_2$ leads to 1 in 4 Ce moments no longer being shielded, then this does not hold true when other ions are being substituted. CeRu$_2$Si$_2$ is an interesting example.\cite{knafo} This system is a heavy fermion system that can be driven to a long-range ordered phase by lattice expansion, similar to expanding the CeFe$_2$Ge$_2$ lattice. Thus, CeRu$_2$Si$_2$  is close to a quantum critical point, but on the heavy fermion side. When about 8\% of the Ce ions are substituted\cite{knafo} for non-magnetic La, then the system changes from a heavy fermion phase to an ordered phase [ordering wavevector (0.69,0,0)]. When about 6\% of the Si ions are substituted\cite{sullow} with Ge ions, then the system also enters the ordered phase. Neither substitutions are anywhere close to the percolation limit, and therefore, La or Ge substitution does not result in a one-to-one correspondence with Ce moments being immune to shielding. The Ge substitution is an iso-valent substitution of the nearest neighbors of the Ce-ions, whereas Ru for Fe is an iso-valent substitution as far away from the Ce moments as possible. Thus, if Ge substitution in CeRu$_2$Si$_2$ were to result in cluster formation through the percolation scenario, then this must occur because the Ge substitution affects all four nearest Ce neighbors of a Si ion. We do not know if this is a realistic scenario.

Neutron scattering experiments\cite{knafo} on quantum critical Ce$_{0.925}$La$_{0.075}$Ru$_2$Si$_2$ have unearthed some tantalizing hints that this system might harbor clusters and CeRu$_2$Si$_2$ might be similar to underdoped Ce(Fe$_{1-x}$Ru$_x$)$_2$Ge$_2$, such as for the Ru concentration x= 0.13 discussed in this paper (data shown in Fig. \ref{in6next}). Inelastic neutron scattering has shown\cite{knafo} that in quantum critical Ce$_{0.925}$La$_{0.075}$Ru$_2$Si$_2$ the magnetic scattering signal increases with decreasing temperature at the ordering wavevector and E= 0 (Fig. 1 in Ref. [\onlinecite{knafo}]), as expected for the onset of ordering. Away from the ordering wavevector, the data showed an increase in scattering, but this increase saturated below T= 5 K (Fig. 2 in Ref. [\onlinecite{knafo}]).  Moreover, the data at the ordering wavevector and away from the ordering wavevector were identical for energy transfers E $>$ 3 meV. In the cluster scenario, such behavior would be associated with smaller clusters forming first, and persisting down to the lowest temperatures, with the largest clusters only emerging close to the QCP. One additional intriguing observation is that $\mu$SR measurements have revealed\cite{longrange} the existence of (small) ordered moments in the disordered, stoichiometric phase. However, without magnetic correlation length data along different crystallographic directions all cluster thoughts remain speculative.

Our final remark is that substitution is not the only way in which clusters could form. Substitution leads to differences in atomic separations of $\sim$ 0.01 nm. This size of displacement also occurs naturally\cite{phonon} in all systems (including stoichiometric systems) when phonons pass through the system. Even at zero Kelvin, the lattice is populated by a considerable number of phonons, and therefore, if we were to take a snapshot at any particular time we would see a distribution of atomic separations. This distribution leads to a distribution of Kondo shielding temperatures. These distributions would change on a ps time scale. However, provided these time scales are much slower than the electronic time scales, then from an electronic point of view there is no essential difference between a distribution originating from doping or one from phonons. We would observe cluster formation in both cases, albeit that in the phonon case these clusters would be fleeting. We are currently investigating this avenue. 

\section{Acknowledgment and Disclaimer}
Part of this material (HB3 and DCS experiments) was based upon work supported by the
Department of Energy under Award Number DE-FG02-07ER46381. The identification of the equipment used in the various measurements is not intended to imply recommendation or endorsement by the National Institute of Standards and Technology, nor is it intended to imply that this equipment is necessarily the best available for the purpose. 

\appendix*
\section{Appendix: Experimental Methods}
\subsection{Sample growth and characterization}
Most of the results in this paper pertain to a single crystal of   Ce(Fe$_{0.76}$Ru$_{0.24}$)$_2$Ge$_2$ that 
was grown\cite{nale} using the floating zone furnace technique. The original, single crystal of about 5 cm in length and 7 mm in diameter is shown in Fig. \ref{sxtal}. The crystal, grown from starting materials Ce (3N5), Ru (3N5), Ge (5N), Fe
(4N8) displayed a small concentration gradient along the cylindrical axis. The sample composition was determined at the top and at the bottom, using an electron
probe microanalyzer (EPMA) JEOL JXA-8621. We found the compositions to be mainly
stoichiometric with a few percentages of Ge-rich secondary phases. The pictures of the top and bottom (see Fig.
\ref{secphase} for pictures of the bottom) show the secondary phases as white stripes. Both ends of the crystal yielded very
similar pictures. The sample composition (normalized to 5 atoms per formula unit)
was found to be Ce$_{0.988}$(Ru$_{0.233}$Fe$_{0.777}$)$_2$Ge$_{1.993}$ for the top part of
the sample and Ce$_{0.996}$(Ru$_{0.261}$Fe$_{0.753}$)$_2$Ge$_{1.978}$ for the bottom part. Bearing this range in mind, we refer to our crystal as having the nominal composition   Ce(Fe$_{0.76}$Ru$_{0.24}$)$_2$Ge$_2$. Based on the phase
diagram (Fig. \ref{fig2}), the top is paramagnetic while the bottom is (just) in the ordered phase,
consistent with the neutron scattering findings (see Fig. \ref{lro}). The [110] crystallographic direction makes an angle of 5 degrees with the cylindrical sample axis.

Once it was discovered in neutron scattering experiments that a resolution-limited component appeared in elastic scans below T= 2 K (see Fig. \ref{lro}), indicative of long-range magnetic order consistent with a ruthenium concentration just in the magnetically ordered part of the phase diagram,
the bottom 1 cm of the crystal was masked in
the initial HB3 neutron scattering experiments, and cut off for subsequent HB3,
DCS, TRIAX, and BT7 experiments. The remainder 11 g crystal did not show Bragg peaks indicative of long-range order or 
polycrystallinity (in neutron scattering experiments or in Laue backscattering). A piece cut off from the top was used for the characterization, specific heat, resistivity, and
susceptibility measurements described in this paper. 
\subsection{Transport measurements}
\begin{figure}
\begin{center}  
\includegraphics*[viewport=40 100 600 550,width=85mm,clip]{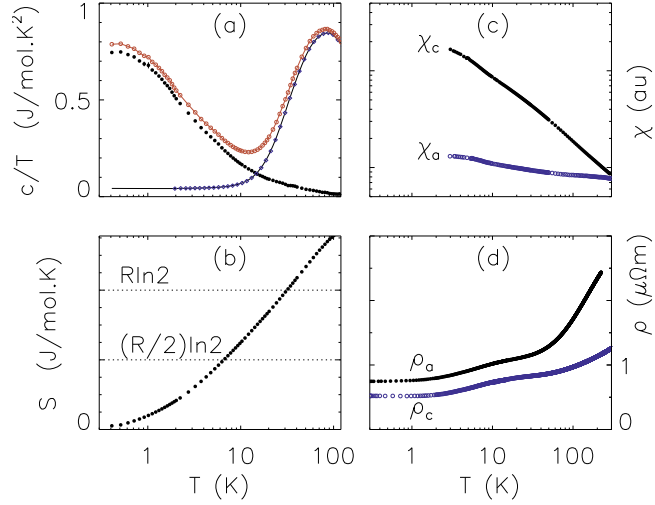}
\end{center}
\caption{(color online) Specific heat (a), entropy (b), susceptibility (c), and resistivity (d) for Ce(Fe$_{0.76}$Ru$_{24}$)$_2$Ge$_2$ measured on a piece of the single crystal used in  our scattering experiments. (a) The specific heat data (open circles, top curve) were corrected for the specific heat of non-magnetic LaFe$_2$Ge$_2$ (diamonds) to yield the magnetic specific heat (filled circles) and are plotted as c/T. The  LaFe$_2$Ge$_2$ data were extrapolated from 2 K down to 0.4 K (solid line). Note the logarithmic horizontal temperature scale. (b) The c/T data were integrated numerically to yield the molar entropy S. R stands for the gas constant of 8.31 J/mol.K. (c) The uniform susceptibility $\chi$ measured along the c-axis (easy axis, top curve) and along the a-direction (hard direction, bottom curve). Both the horizontal and vertical axis are logarithmic axes. (d) The resistivity $\rho$ as measured along the a-direction (top curve) and the c-direction (bottom curve).} 
\label{fig3}
\end{figure}

\begin{figure}
\begin{center}  
\includegraphics*[viewport=100 125 530 680,width=85mm,clip]{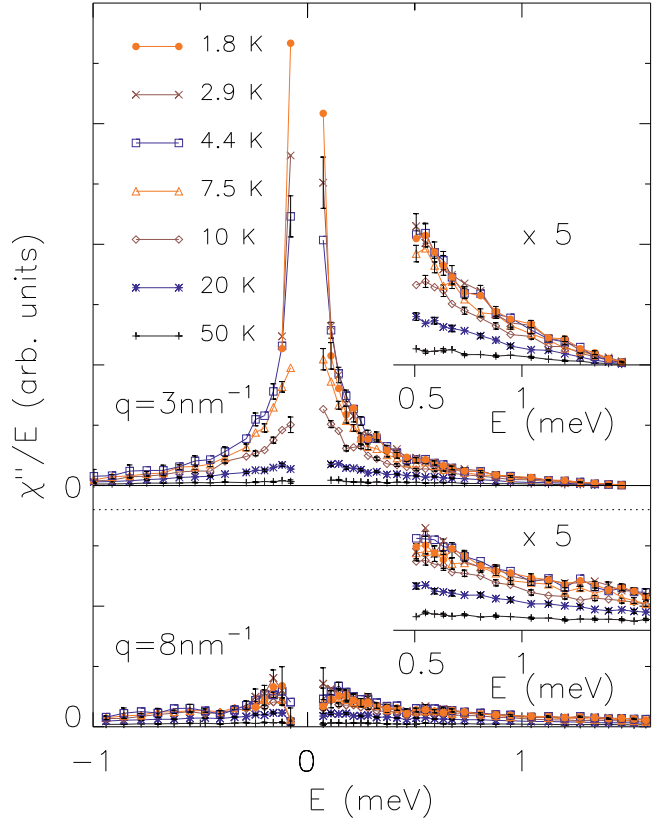}
\end{center}
\caption{(Color online) Constant q-cuts through IN6 data on polycrystalline Ce(Fe$_{0.76}$Ru$_{0.24}$)$_2$Ge$_2$ at T= 1.85, 2.86, 4.36, 7.5, 10.4, 19.6, and 49.8 K as a function of energy E transferred from the neutron to the sample. The data were corrected according to Ref. [\onlinecite{wouterprl}] and have been plotted after the Bose population factor has been taken out: $\chi"(q,E)/E= (1-e^{-E/k_BT})S(q,E)/E$. The energy resolution of the spectrometer was 0.07 meV; data within this window have not been plotted as the incoherent elastic scattering could not be sufficiently corrected for. The data at the ordering wavenumber q= $|\vec{Q}|$= 3 $\pm$ 0.1 nm$^{-1}$ (above dotted line) display the onset of ordering upon cooling, with the additional scattering (upon cooling) exhibiting an increasingly narrow width in energy. The higher energy transfers are shown on an enlarged scale. Note that, in this energy range, there is very little temperature dependence to the data for T $<$ 5 K. The lower data set (below dotted line) is a similar cut (q= 8 $\pm$ 0.1 nm$^{-1}$; the data sets share the same vertical scale) but taken away from the ordering wavenumber. The symbols are denoted in the figure. For plotting clarity, only every other error bar has been shown.} 
\label{ill}
\end{figure}

\begin{figure}
\begin{center}  
\includegraphics*[viewport=100 110 480 600,width=85mm,clip]{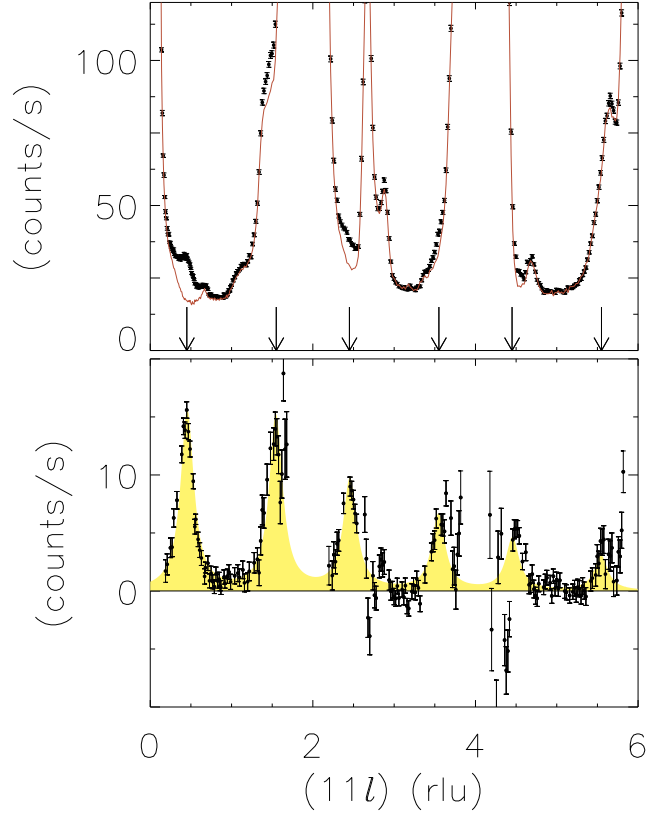}
\end{center}
\caption{(Color online) The top panel displays the elastic HB3 scattering data on an 11 g single crystal of Ce(Fe$_{0.76}$Ru$_{24}$)$_2$Ge$_2$. There is a clear difference between the data measured at 2 K (points) and at 56 K (red line). Subtracting these two data sets shows the incipient order that develops (bottom panel; adapted from \protect\cite{wouterprb}) at the ordering wavevector $\vec{Q}$= (0,0,0.45) as indicated by the peaks at (1,1,2n $\pm$0.45) with n an integer (shown by arrows). Note that this signal is difficult to observe on top of the large nuclear background consisting of nuclear Bragg peaks at (1,1,2n) and alumininum powder lines. The decrease in magnetic scattering with increased momentum transfer follows the cerium form factor \protect\cite{wouterprb}. The filled areas under the data in the lower panel are given by a Lorentzian-line fit to the data\protect\cite{wouterprb}.} 
\label{sro}
\end{figure}

\begin{figure}
\begin{center}  
\includegraphics*[viewport=20 120 490 560,width=85mm,clip]{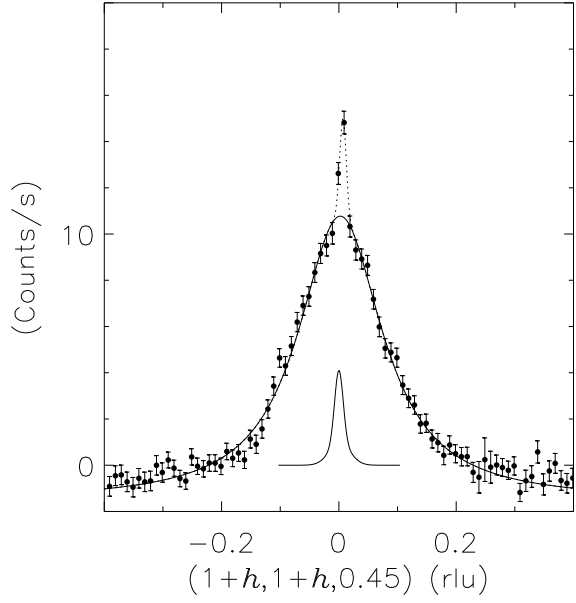}
\end{center}
\caption{When the lower 1 cm Ru-rich part of our single crystal of Ce(Fe$_{0.76}$Ru$_{24}$)$_2$Ge$_2$ is not shielded (or cut off) in elastic scattering experiments, then long-range order can be seen to develop in addition to the short-range order present  in the quantum critical part of the crystal. The long-range order (at the same ordering wavevector) is seen as a resolution limited Bragg peak (dotted line) on top of the broad short-range order that was visible in Fig. \ref{sro} with the lower part of the crystal masked. These data were taken on HB3, and the difference signal between 2 K and 56 K is displayed in the figure. The negative intensities away from the ordering wavevector are caused by the diminished scattering at these wave numbers owing to Kondo shielding and a transfer of magnetic intensity from the broad background to the short-range ordered signal. The solid line through the points is a Lorentzian with full width at half maximum of 0.166 r.l.u. The stand-alone solid line is a line through the data points for the (1,1,0) nuclear Bragg peak which is shown on a different intensity scale; this Bragg peak serves as the resolution linewidth along the longitudinal direction.} 
\label{lro}
\end{figure}

\begin{figure}
\begin{center}  
\includegraphics*[viewport=50 90 580 440,width=85mm,clip]{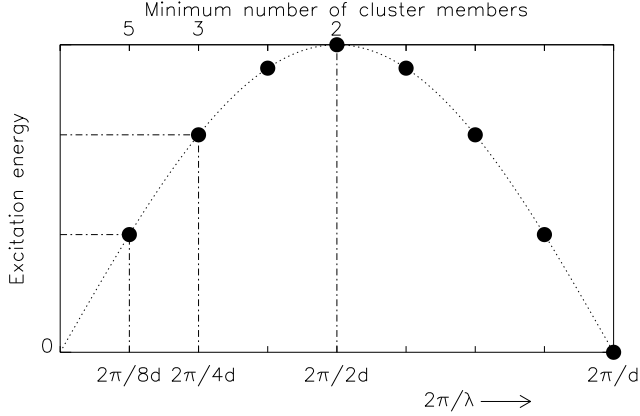}
\end{center}
\caption{Finite-size effects force the moments on an isolated cluster to line up at low temperatures. The dotted sinusoidal curve is a typical dispersion\protect\cite{nikotin} for an antiferromagnetic system, displaying the energy cost (vertical axis) required to impose a disordering disturbance of wavelength $\lambda$ (horizontal axis). In here, d stands for the separation between magnetic moments. Typical energy costs \protect\cite{nikotin} of disturbances of wavelength $\lambda$= 2d, the equivalent of neighboring spins being misaligned, are of the order of a few hundred to a few thousand Kelvin. A disordering fluctuation on a cluster has the additional quantum mechanical requirement that the wavelength has to be a half-integer times the cluster diameter D: n$\lambda$= 2D. This requirement breaks up the continuous dispersion into a collection of points.\protect\cite{orbach} For instance, the disordering fluctuation with the longest wavelength (lowest energy) permissable on a cluster of linear size 4d is $\lambda$= 8d. This example is given by the left-most solid black dot, and its energy cost is displayed by the horizontal dashed-dotted line. The permitted wavelengths for other fluctuations on this particular cluster are shown by the other black dots. Note that a wavelength of $\lambda$= d corresponds to a fluctuation of zero energy cost since the phase difference between the moments is 360$^o$. Thus, in order for the moments to misalign, a finite energy cost is required, dictated by the size of the cluster. If the thermal energy available is considerably less than this energy cost then clusters will be fully ordered.} 
\label{finite}
\end{figure}

\begin{figure}
\begin{center}  
\includegraphics*[viewport=10 70 560 520,width=85mm,clip]{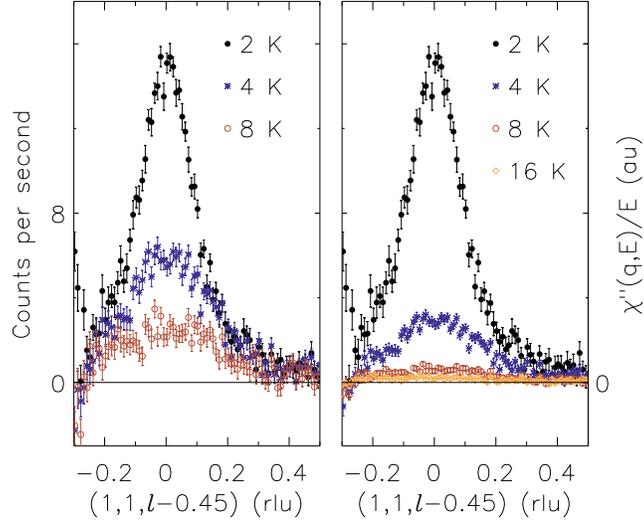}
\end{center}
\caption{(Color online) Elastic scattering data obtained by subtracting the scattering measured at the temperatures shown in the figure from the data at 56 K. These data, measured on HB3 as a function of q= (1,1,{\it l}) are given as the dynamic structure factor S(q,E) in the left panel, and as the reduced imaginary part of the susceptibility $\chi$"(q,E)/E= (1-e$^{- E/k_BT}$) S(q,E)/E in the right panel. For elastic scattering, this conversion amounts to a division by the temperature. Note that the tails of the nuclear Bragg peak at (1,1,0) influence and start to overwhelm the signal below {\it l}= 0.2 (or, {\it l} - 0.45 $<$ -0.2).} 
\label{additional}
\end{figure}

The resistance, specific heat, and susceptibility measurements were performed on two
pieces of the sample of sizes 1.2mm x 0.85mm x 6.0 mm, weighing 50 mg, taken from
the top part of the sample. The resistance data  (shown in lower right panel of Fig. \ref{fig3}) were taken using an Oxford Instruments
MagLab for the temperature range 2K - 300K, while an
Oxford Instruments Heliox VL was used for the range 0.3K - 15K. The electrical contacts were placed in the standard 4-probe geometry, using Cu wires and
silver paste. The ac-currents were along [100] (upper curve in Fig. \ref{fig3}, $\rho_0$= 74.3 $\mu \Omega$cm)
and [001] (lower curve, $\rho_0$ $\approx$ 52 $\mu \Omega$cm). While the room temperature resistivity values
are similar to those measured\cite{cefege} in pure CeFe$_2$Ge$_2$ (250 and 85  $\mu \Omega$cm along [100] and
[001], respectively), the  $\rho_0$ values are considerably larger, reflecting disorder attributable to Ru
substitution on Fe-sites.

The resistivity data show the onset of coherence around T $\sim$ 15 K. The resistivity data taken with
the current along [001] bottom out at T= 0.8 K, below which they show a small increase.
The data along [100] do not show this increase. The [100]-data vary with temperature as
T$^{1.5}$ in the range 1.5 K $\le$ T $\le$ 3.5 K, with a higher exponent (1.86) below 1.5 K, and a
lower exponent above 3.5 K.

We measured the specific heat on one of the 50 mg pieces cut from the top of the crystal.
The data were collected using Quantum Design Physical Properties Measurement
Systems, while the magnetic contribution was isolated by subtracting the specific heat values
for a non-magnetic isostructural polycrystalline reference sample (LaFe$_2$Ge$_2$, top left panel of Fig. \ref{fig3}). The data
show non-Fermi liquid behavior over a large temperature range, but similar to the
results for the resistivity, we find at the lowest temperatures that the data show signs of
the sample being slightly on the paramagnetic side of the QCP. The coefficient of the
linear term in the specific heat at T= 0.3 K is $\gamma$= 748 $\pm$ 5 mJ/mol.K$^2$.

The susceptibility was determined using a Quantum Design SQUID magnetometer in the
temperature range 3.5 K - 300 K. We used two 50 mg samples, taken from the top part of
the crystal. We collected data in a field of 0.2 T (the magnetization was found to be linear
in fields up to 0.5 T). The data shown in thetop right panel Fig. \ref{lro} confirm that the [001]-axis is the easy
axis.
\subsection{Neutron scattering experiments}
We performed new neutron scattering experiments at the BT7 polarized neutron spectrometer\cite{bt7ii} at the National Institute of Standards and Technology (NIST), and at the TRIAX triple-axis spectrometer at the University of Missouri Research Reactor (MURR). Other neutron scattering experiments revisited and written up in this paper have been described elsewhere.\cite{wouterprl,wouterprb} The BT7 spectrometer was operated in full polarization mode,\cite{bt7} utilizing $^3$He polarizers in the incident and scattered beam. The spectrometer was operated at a fixed final energy of 14.7 meV, while second and third order neutrons were filtered out before reaching the detector by a PG-filter in the scattered beam. This particular set up, including the collimators used to limit the angular divergence of the incoming and scattered neutrons, resulted in an energy resolution (full width at half maximum) of 0.8 meV. The polarizers were exchanged on a daily basis during the 5 day long experiment; the time-dependence of each polarizer employed was measured repeatedly during the course of the experiment, allowing for a separation between the spin-flip, and non spin-flip channels by using the software developed\cite{bt7} for this purpose at NIST. The term spin-flip indicates that the neutron underwent a change in its magnetic moment direction, while non spin-flip indicates that the neutron had the same moment orientation before and after interacting with the sample.

The scattering cross-section for magnetic interactions between a neutron and the sample depends both on the relative orientations of the neutron and cerium moments, as well as on the direction of the momentum transferred by the neutron to the sample.\cite{scattering,polscat} For the experiments we oriented our sample with the [110] and the [001] direction in the scattering plane, so that momentum could be transferred along these crystallographic directions, or to combinations of these directions. The magnetic cross-section is such that when the neutron polarization direction is chosen to be parallel to the direction of momentum transfer, then the only scattering that shows up in the spin-flip channel is scattering that is magnetic in origin.\cite{polscat} This is the set up we chose for all polarized neutron scattering experiments discussed in this paper.
 
The (magnetic) signal to noise ratio of a polarized scattering experiment allows for the measurement of a weak magnetic signal in the presence of non-magnetic scattering. However, the measured spin-flip is not completely free from unwanted counts: there are some neutrons that reach the detector from other directions, and some spurions (neutrons that reach the detector by means of multiple, or unusual scattering processes) can also be present. The largest source of unwanted scattering in the spin-flip channel is magnetic incoherent scattering attributable to the nuclear spins of the various ions in the sample. For this reason, and as a check on our data reduction procedure, we compare our low-temperature data to higher temperature data; the reason behind this is that practically all of the unwanted scattering, magnetic incoherent scattering included, does not depend on the sample temperature, and as such, a direct comparison between low and high temperature data will result in a signal that is purely magnetic in origin.

The TRIAX set up and data analysis is much more straightforward. TRIAX was operated at a fixed final energy of 14.7 meV with PG-filters in the scattered beam. The energy resolution was similar to the resolution of BT7. The data reduction is simply a matter of counting it out at low and higher temperatures, and use the difference signal for the interpretation of the data. This procedure is effective because the BT7 experiments identified regions in reciprocal space (suitable to non-polarized experiments) that are largely free of unwanted nuclear scatttering, such as the temperature dependent phonon scattering that tends to obfuscate the information hidden in the difference signal in non-polarized neutron scattering experiments.

\begin{figure}
\begin{center}  
\includegraphics*[viewport=35 96 480 630,width=85mm,clip]{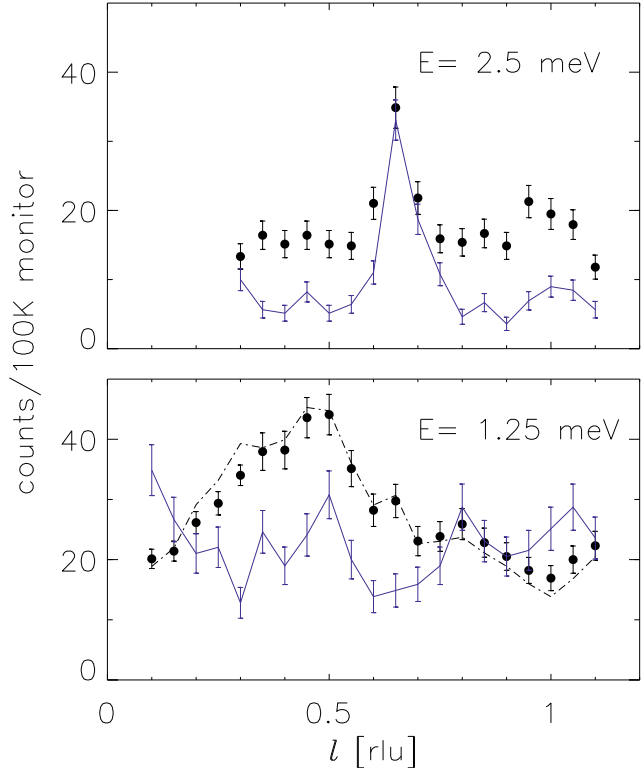}
\end{center}
\caption{ (Color online) Typical raw data for our fully polarized BT7 experiment measured along (1,1,$l$) for the fixed energy transfers shown in the panels. The filled circles with error bars are the data measured in the spin-flip channel, the solid lines with error bars are the data in the flipper off channel. The flipper-on data in the top channel were taken at T= 50 K, while the nuclear, flipper-off data are the sum over two datasets taken at 5 K and 50 K. Since the latter two data sets did not differ outside of the error margins, we summed them and normalized to the monitor count. The top panel demonstrates that broad magnetic scattering is present at T= 50 K. The bottom panel shows the magnetic scattering at 1.5 K, and the nuclear scattering at 5 K. When the magnetic scattering is corrected for the non-ideal and time-dependent polarization efficiency\protect\cite{bt7} of the two helium-3 polarizers, then the magnetic data are given by the dashed-dotted line. The small difference between the corrected and uncorrected magnetic data demonstrates that our magnetic signal is largely free from unwanted nuclear scattering at E= 1.25 meV. The peak at $l$= 0.67 r.l.u. in the top panel is a spurion (see Appendix).} 
\label{allmagnetic}
\end{figure}

\begin{figure}
\begin{center}
\includegraphics*[viewport=50 80 460 630,width=85mm,clip]{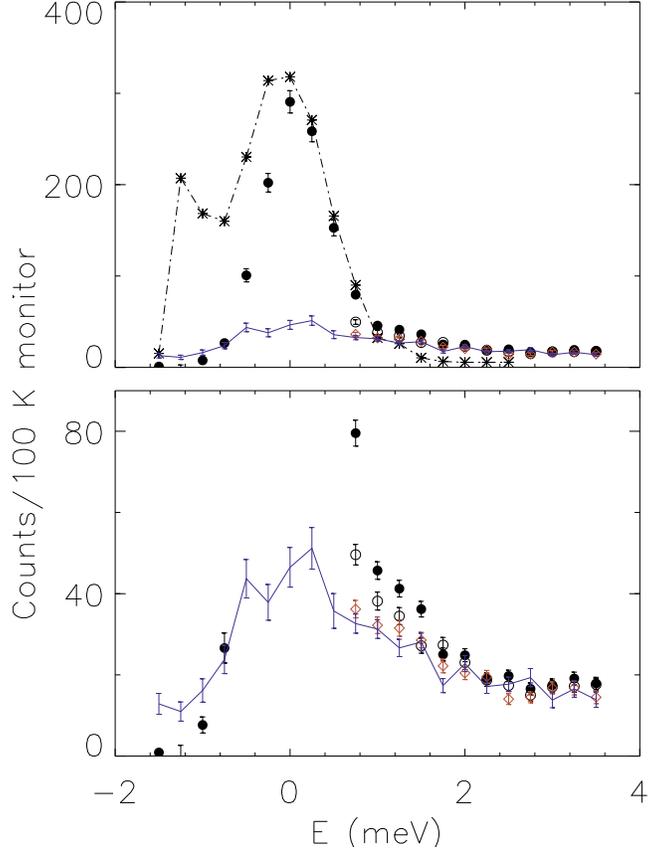}
\end{center}
\caption{(Color online) Both panels display the energy E and temperature dependence of the scattering measured at $\vec{Q}$= (1,1,0.45) using the BT7 spectrometer in full polarization mode. The bottom panel is an enhanced vertical scale of the top panel. The filled circles are data measured in the spin flip channel at 1.56 K, the open circles at 5 K, the open diamonds at 20 K, and the line with error bars are the data at 50 K. The dashed-dotted line with the asterisks symbols are the non-spin flip scattering data taken at 1.56 K. This latter data set is used in the correction procedure to correct for the seep-through of the nuclear scattering into the magnetic spin-flip scattering channel. The messy nuclear data serve as a reminder of how difficult it is to measure the very weak magnetic ordering signal. The peak at E= -1.5 meV in the nuclear data results in a slight overcorrecting of the magnetic signal when doing the polarization correction. The bottom panel shows that the magnetic scattering increases for all energy transfers E $>$ 0 meV with the caveat that the error bars are too large for E $>$ 2 meV to draw detailed conclusions.} 
\label{quasi}
\end{figure}

\begin{figure}
\begin{center}
\includegraphics*[viewport=40 80 550 620,width=85mm,clip]{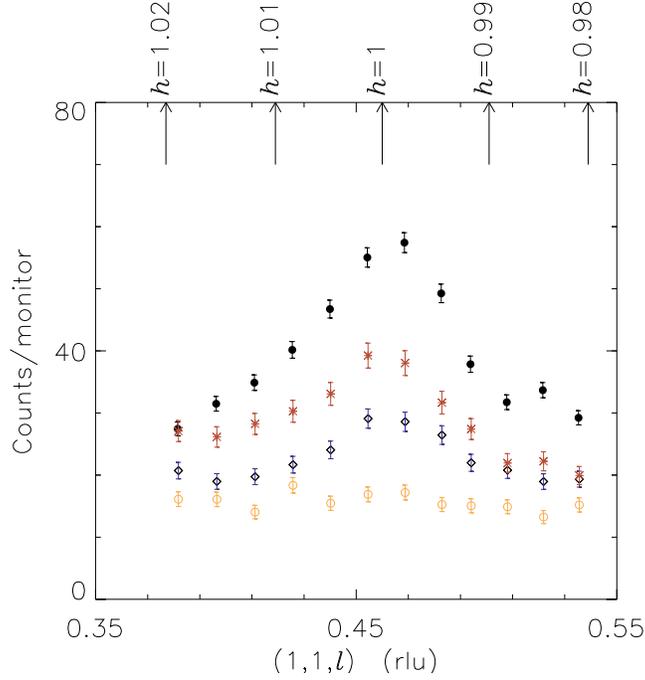}
\end{center}
\caption{(Color online) The temperature dependent elastic scattering near $\vec{Q}$= (1,1,0.45) as measured on DCS. The filled circles are measured at 0.4 K (top set of points), the stars at 0.9 K, the diamonds at 1.8 K and the open circles (bottom set of points) at 56 K. The data are the raw data normalized to the incident beam monitor and summed over the energy range -0.075 $<$ E $<$ 0.075 meV. The {\it l}$-$values are shown on the horizontal axis, while the {\it h}$-$ and {\it k}$-$values are given at the top of the figure. The data show that the scattering at increasingly lower temperatures appears above the already existing scattering levels. Note that because the amount of momentum transferred along all crystallographic axes varies for each data point, we cannot conclude anything about the lineshape other than that magnetic Bragg peaks do not appear down to 0.4 K, indicating that the sample remains in the (nominally) paramagnetic phase.} 
\label{dcsord}
\end{figure}

\begin{figure}
\begin{center}
\includegraphics*[viewport=40 80 580 620,width=85mm,clip]{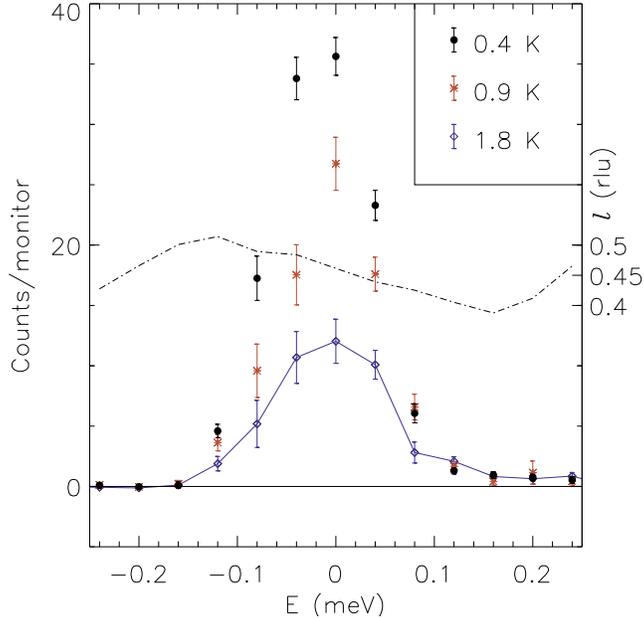}
\end{center}
\caption{(Color online) The net scattering data measured on DCS at the ordering wavevector  $\vec{Q}$= (1,1,0.45)  are shown for three temperatures (indicated in the figure). The data in the figure were obtained by grouping the scattering data in energy bins 0.04 meV wide (taking the center of the bin as the energy values displayed in the figure). The binned data were then summed over 0.99 r.l.u. $<$ {\it h,k} $<$ 1.01 r.l.u., and the signal at 56 K was subtracted from all three datasets shown in the figure. Since $\vec{q}$ and E vary for each data point, the value of {\it l} displays an energy dependence. This energy dependence is shown by the dashed-dotted line and the vertical scale on the right hand side. The line drawn through the T= 1.8 K points simply connects the data points. It can be seen from the figure that the additional scattering that develops below 1 K is restricted to energies E with $|$E$|<$ 0.1 meV. The data are slightly asymmetric around E= 0 because of the asymmetric Ikeda-Carpenter type resolution function\protect\cite{ik} characteristic of time-of-flight spectrometers, moving the scattering to lower energy transfers and producing a negative energy tail. The characteristic half width at half height of the scattering at 0.4 K is 0.05 meV, which is the energy resolution of the DCS spectrometer. Note that the scattering for E $>$ 0.1 meV is significantly different from zero (implying there is more scattering at these temperatures than at T= 56 K), and that this scattering is virtually independent of temperature, suggesting that clusters that form below 2 K do not fluctuate on the time scales less than 80 ps.} 
\label{dcsenergy}
\end{figure}

\begin{figure}
\begin{center}
\includegraphics*[viewport=100 110 500 550,width=85mm,clip]{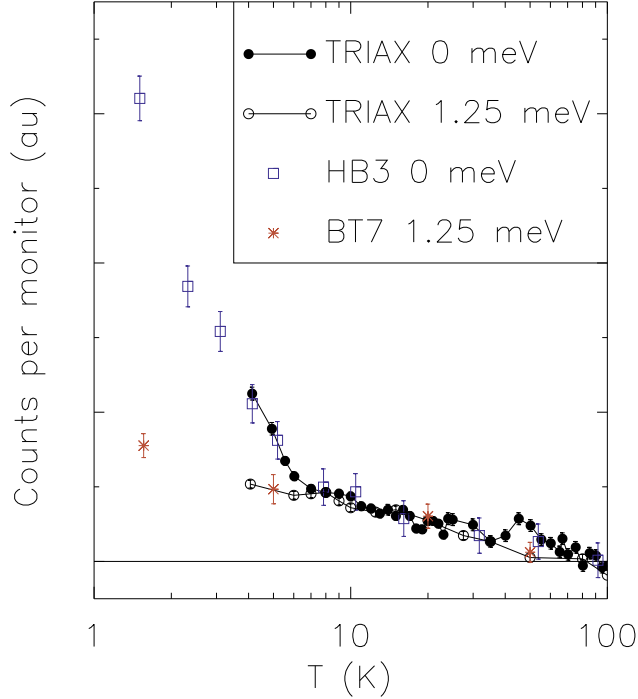}    
\end{center}
\caption{(Color online) The temperature dependence of the scattering at $\vec{Q}$= (1,1,0.45) as measured on TRIAX, HB3, and BT7 for the elastic channel E= 0 meV, and for the inelastic channel at E= 1.25 meV. For all instruments, the latter energy transfer is sufficiently far removed from E= 0 that resolution broadened elastic scattering is not measured. The vertical scale is different between all 4 datasets. We subtracted the scattering at the highest temperature so that the solid horizontal line measures the scattering with respect to the level at 100 K. We then placed the elastic HB3 and TRIAX scattering data on the same level by multiplying the TRIAX scattering by a constant factor, chosen the achieve the best agreement between the two datasets over as large a temperature range as possible. This establishes the temperature dependence of the elastic scattering. Note that the TRIAX experiment was set up to measure this critical scattering with high accuracy, as opposed to the HB3 experiment that had a different primary purpose. The same holds true for the inelastic data where the TRIAX data are more accurate than the BT7 data, but the BT7 data extended to lower temperatures. The inelastic data were put on the same scale the same way as the elastic data, and in addition, the inelastic datasets were multiplied by a constant factor so that the temperature dependence of the elastic and inelastic data would be similar over the largest range possible. The data show that the temperature dependence of the elastic and inelastic data are qualitatively different, with the differences appearing below T $<$ 10 K.}
\label{ordscat}
\end{figure}

\begin{figure}
\begin{center}  
\includegraphics*[viewport=50 100 560 600,width=85mm,clip]{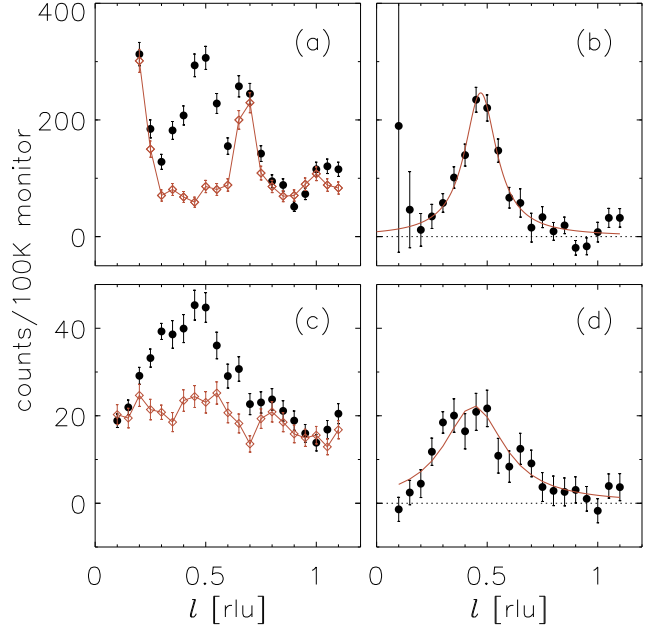}
\end{center} 
\caption{(Color online) The width in reciprocal space for short-range order scattering at E= 0 meV (top panels) and at E= 1.25 meV (bottom panels) as measured on BT7. Panels (a) and (c) display the spin-flip scattering data measured along (1,1,$l$) at 1.56 K (filled symbols) and at 50 K (open symbols with line connecting the points). The difference between the low and high temperatures are shown in panels (b) and (d) as filled symbols. The solid lines in panels (b) and (d) are best fits to a Lorentzian line with full width at half maximum of (0.17 $\pm$ 0.02) r.l.u. (panel (b) for E= 0 meV) and (0.33 $\pm$ 0.05) r.l.u. (panel (d) for E= 1.25 meV). See Appendix for a discussion on the spurious bump at $l$= 2/3 r.l.u. The data at the lowest $l$ in panel (a) show the seep-through of the strong nuclear Bragg peak at (1,1,0). While the polarization correction procedure removes most of the nuclear scattering, it is not 100 \% effective. Similarly, the data at T= 50 K in panel (c) at the lowest $l$ values appear to show some temperature dependent nuclear scattering (phonons). As such, the lowest $l$ data should be taken with a grain of salt, representing the limits of what polarizers can correct for when flipper off data are not taken at each and every corresponding temperature because of beam time restrictions.} 
\label{inelcorlength}
\end{figure}

\begin{figure}
\begin{center}  
\includegraphics*[viewport=50 120 500 650,width=85mm,clip]{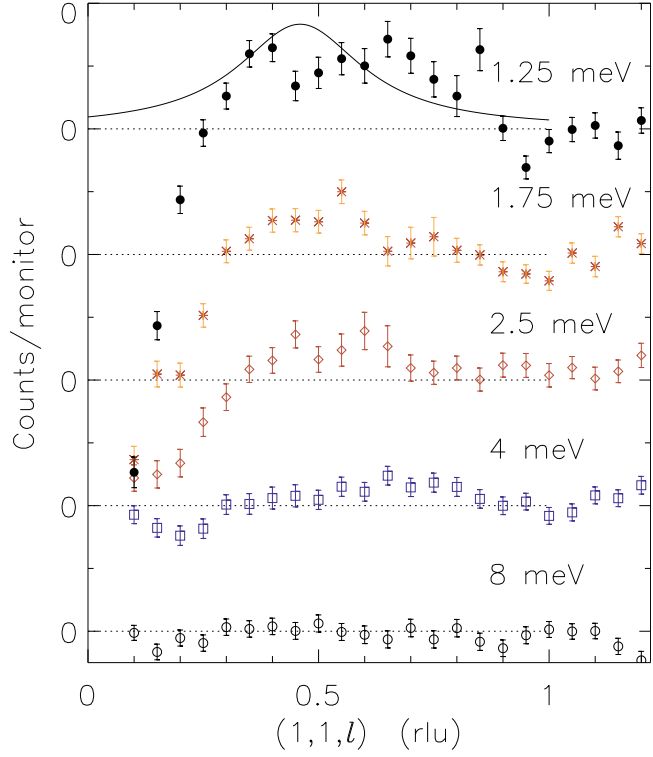}
\end{center}
\caption{(Color online) The net neutron scattering counts at 4.5 K as measured on TRIAX after subtraction of the counts at 80 K. The data are obtained as a function of momentum transfer along (1,1,{\it l}) for fixed energies shown in the figure. Each energy panel is offset along the vertical axis for clarity, with the dotted lines representing zero net counts. While the data show that there is more scattering at 4.5 K than at 80 K around the ordering wavevector for energy transfers up to 4 meV, the data are too noisy to determine the width in reciprocal space. The solid line in the 1.25 meV data panel is the best fit from the NIST data at the same energy and 1.56 K (see Fig. \ref{inelcorlength}). The main problem in doing the temperature based subtraction in unpolarized experiments  is the temperature dependence of the phonon scattering associated with the (1,1,0) nuclear Bragg peak. This problem manifests itself as a negative phonon peak located at (q,E)$-$points determined by the phonon dispersion.} 
\label{failed}
\end{figure}

\begin{figure}
\begin{center}  
\includegraphics*[viewport=80 110 500 550,width=85mm,clip]{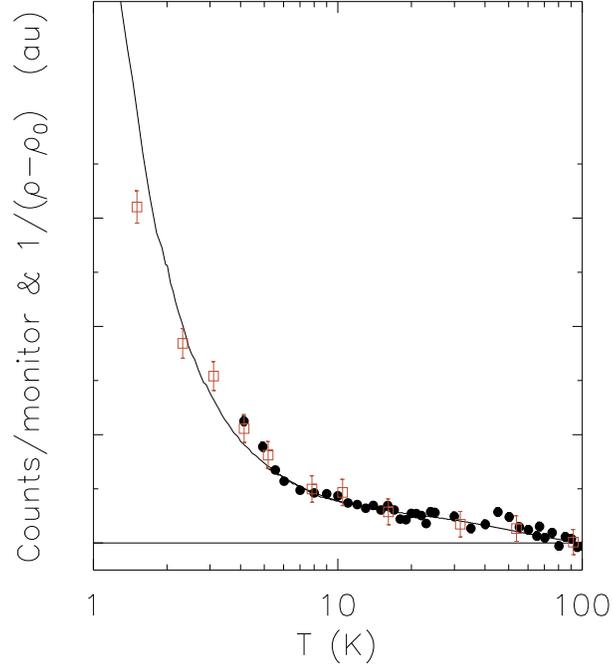}
\end{center}
\caption{(Color online) Comparison between the resistivity and the elastic scattering at the ordering wavevector as measured on HB3 (squares) and TRIAX (filled circles). The neutron scattering data are plotted with respect to their level at 100 K (solid horizontal line) and the TRIAX data were multipled by a constant factor identical to Fig. \ref{ordscat}. The solid line through the points is the measured conductance along the (1,1,0) direction plotted as 1/($\rho$-$\rho_0$) with respect to the value at 100 K of this quantity. These data were then multiplied with a constant factor chosen to achieve the visually best correspondence with the neutron scattering data. The temperature is displayed on a logarithmic scale.} \label{resis}
\end{figure}

\begin{figure}
\begin{center}  
\includegraphics*[viewport=0 0 350 490,width=60mm,clip]{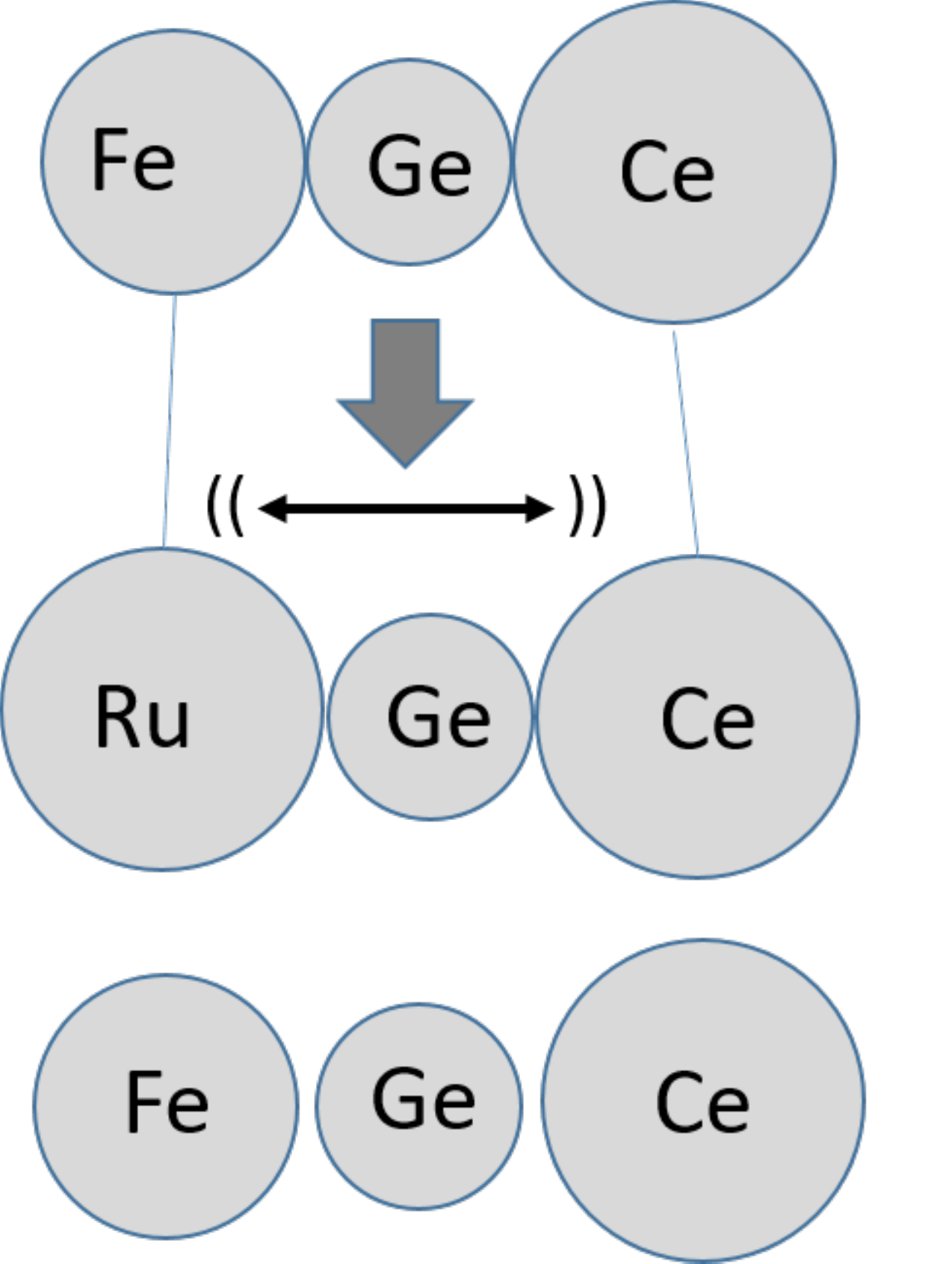}
\end{center}
\caption{When Fe ions in a Fe-Ge-Ce containing lattice are substituted with larger Ru ions, then this will locally result in lattice expansion and some Ce ions will experience less overlap with their surroundings, thereby maintaining their magnetic moment down to 0 K.}
\label{sub}
\end{figure}

\begin{figure}
\begin{center}  
\includegraphics*[viewport=120 115 680 680,width=85mm,clip]{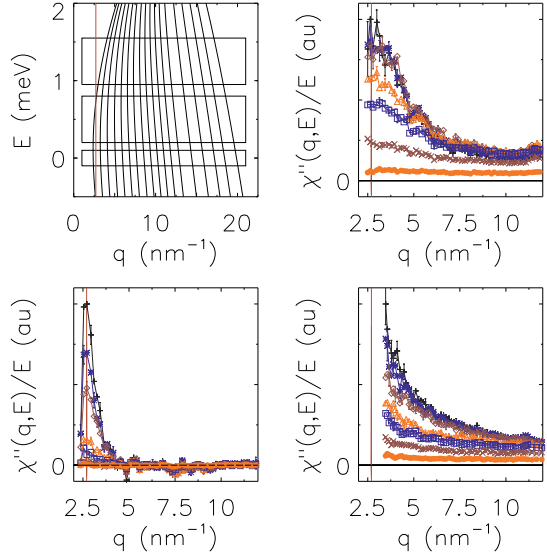}
\end{center}
\caption{(Color online) Cuts through IN6 data on polycrystalline Ce(Fe$_{0.76}$Ru$_{0.24}$)$_2$Ge$_2$ at T= 1.85, 2.86, 4.36, 7.5, 10.4, 19.6, and 49.8 K. The solid lines in the  top left panel show the traces of representative detectors through reciprocal space. The three boxes represent the cuts at constant energy (0 $\pm$ 0.1, 0.5 $\pm$ 0.3, 1.25 $\pm$ 0.3 meV). The results of these cuts are shown in the bottom left, top right and bottom right panels, respectively. The vertical line at q= 2.7 nm$^{-1}$ is given by q=$|\vec{Q}|$. The data were corrected according to the procedure described in ref. \protect\cite{wouterprl}, which includes a direct subtraction of the scattering at 150 K in the range $|$E$| <$ 0.1 meV. The data are shown as $\chi$"(q,E)/E, implying that the Bose population factor has been taken out. The various temperatures are denoted by the symbols '+' (1.85 K), stars (2.86 K), diamonds (4.86 K), triangles (7.5 K), squares (10.4 K),  'x' (19.6 K), and circles (49.8 K). The elastic cut in the lower left panel shows the rapid (with temperature) increase in scattering at the ordering wavenumber. The occasional negative intensities are the result of having subtracted the elastic scattering at 150 K. The cuts at 0.5 and 1.25 meV demonstrate that there is very little temperature evolution below 5 K other than the Bose population changes, and that the characteristic width (in q) of these cuts is significantly larger than the width at E= 0 meV. Note that the three panels do not share an identical vertical scale; only the datasets within the same panel share the same vertical scale. For plotting clarity only every third error bar is shown.}
\label{in6}
\end{figure}

\begin{figure}
\begin{center}  
\includegraphics*[viewport=120 115 680 680,width=85mm,clip]{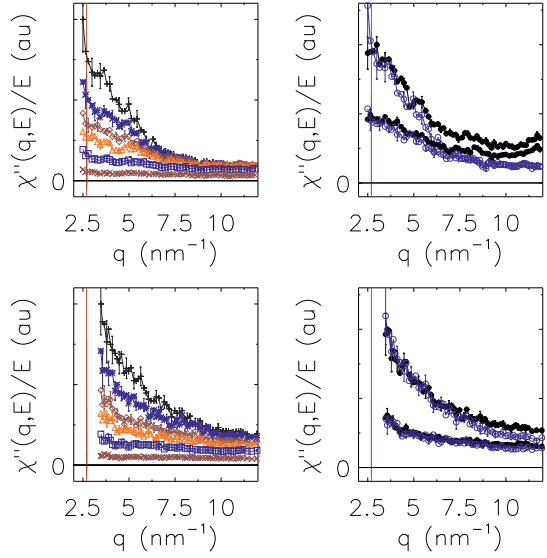}
\end{center}
\caption{(Color online) Cuts through IN6 data on polycrystalline heavy-fermion (HF) CeFe$_{1.7}$Ru$_{0.3}$Ge$_2$ at T= 1.85, 4.38, 7.5, 10.5, 19.7, and 49.8 K (left two panels). The top left panel was taken at E= 0.5 $\pm$ 0.3 meV, the bottom left panel at E= 1.25 $\pm$ 0.3 meV. The cuts were performed identical to the ones on the quantum critical (QCP) sample shown in Fig. \ref{in6}. The various temperatures are denoted by the symbols '+' (1.85 K), stars (4.38 K), diamonds (7.5 K), triangles (10.5 K), squares (19.7 K),  and 'x' (49.8 K). The two panels on the right compare cuts through the quantum critical sample (filled symbols) to cuts through the heavy fermion sample (open symbols). The top right panel shows cuts at E= 0.5 $\pm$ 0.3 meV for T= 1.86 K (HF sample) and 7.5 K (QCP sample) [top two curves] and for T= 7.5 K (HF sample) and T= 15.1 K (QCP sample) [bottom two curves]. The cuts for the different samples at different temperatures demonstrate a large degree of similarity. The lower right panel compares the cuts at  E= 1.25 $\pm$ 0.3 meV for T= 1.86 K (HF sample) and 4.36 K (QCP sample) [top two curves] and for T= 10.5 K (HF sample) and T= 15.1 K (QCP sample) [bottom two curves].  Note that the four panels do not share an identical vertical scale; only the datasets within the same panel share the same vertical scale. For plotting clarity only every third error bar is shown.}
\label{in6next}
\end{figure}

\begin{figure}
\begin{center}
\includegraphics*[width=110mm,clip]{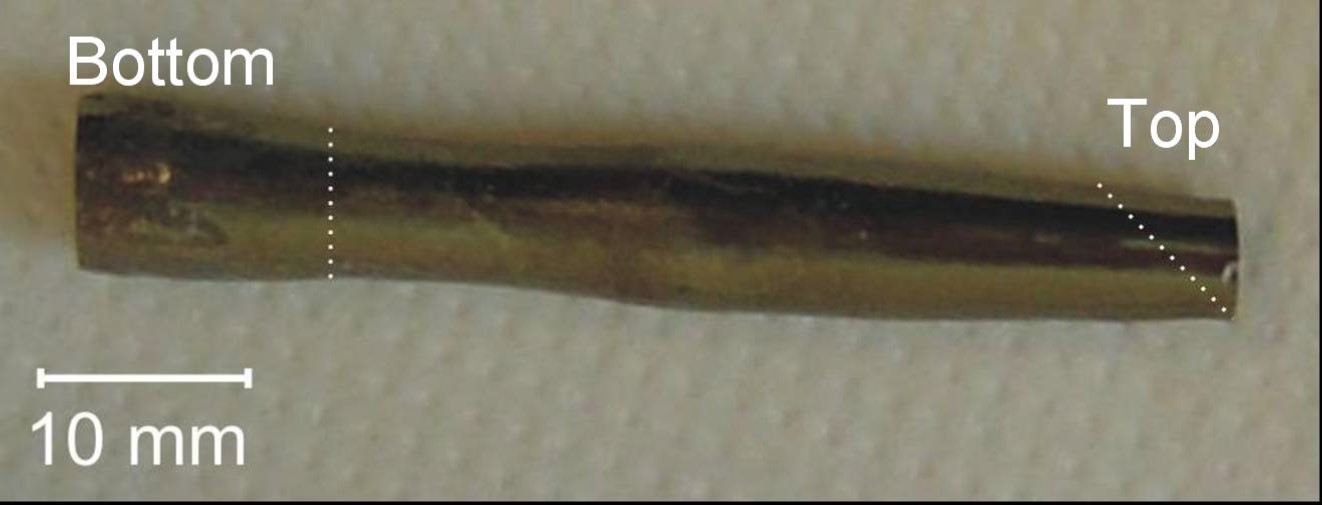} 
\end{center}
\caption{The single crystal \protect\protect\cite{nale} of  Ce(Fe$_{0.76}$Ru$_{0.24}$)$_2$Ge$_2$  used for all single crystal neutron scattering experiments. The part labeled 'bottom' exhibited long-range order (see Fig. \ref{lro}) and was part of the first HB3 experiments, subsequently masked with Gd paint for follow-up HB3 experiments, and cut off in later experiments (DCS, TRIAX, BT7). The part labeled 'top' was cut off and used in specific heat, susceptibility, and resistivity measurements.} \label{sxtal}
\end{figure}

\begin{figure}
\begin{center}  
\includegraphics*[width=110mm,clip]{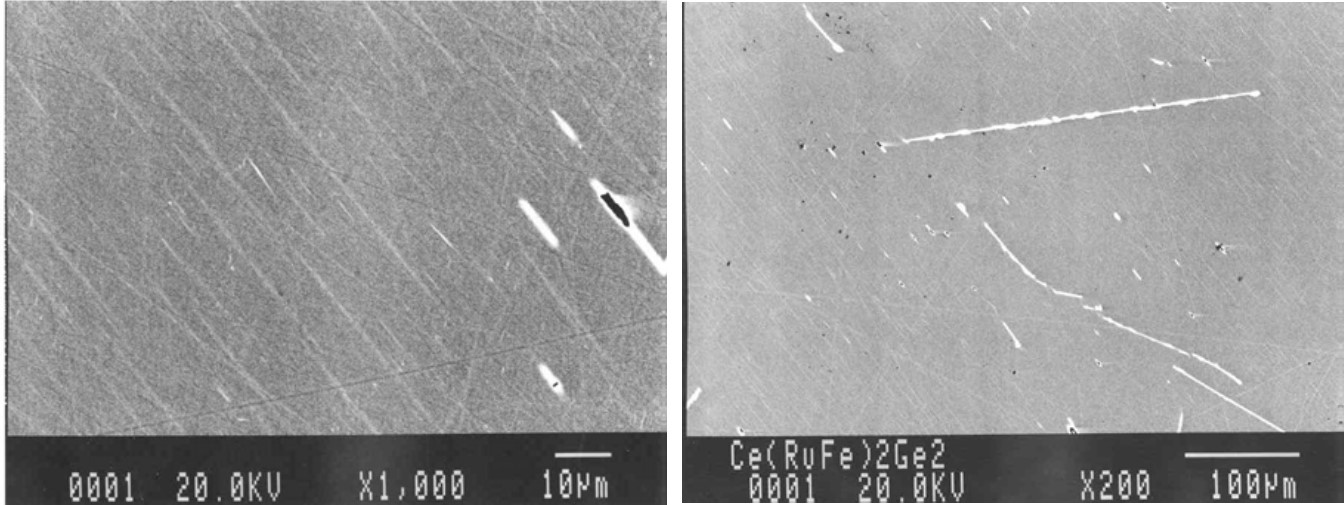}
\end{center}
\caption{Mircroprobe measurements reveal the presence of a Ge-rich secondary phase as white lines. These two photos (showing different magnification) were taken using the bottom section of the crystal.} 
\label{secphase}
\end{figure}

We noted the existence of a spurious signal detected in the inelastic spectrum (Fig. \ref{allmagnetic} at an energy transfer of E= 2.2 meV at q= (1,1,0.67) and (1,1,1.67). This spurion was observed both in the spin-polarized experiment on BT7 and the non-spin-polarized experiment on TRIAX. It is sharply localized in both reciprocal space and energy, with a q-width of 0.01 reciprocal lattice units (r.l.u.) at both positions and energy widths of $\Delta$ E= 1.6 meV at $l$= 0.67 r.l.u. and $\Delta$E= 1.2 meV at $l$= 1.67 r.l.u. as measured on TRIAX. For any true dynamic process, an inelastic scattering signal must obey the detailed balance condition: $S(q,E)= e^{-E/k_BT}S(q,-E)$. With this in mind we measured the signal on Triax at q= (1,1,1.67) and both E= 2meV and E= -2 meV for three different temperatures (5 K, 30 K, and 80 K). These measurements showed that the suprion does not obey detailed balance and is therefore not truly inelastic. We have also characterized the signal using spin-polarization analysis. The non-spin-flip channel was found to have no temperature dependence, whereas the spin-flip channel had an enhanced response for 1.56 K and 5 K compared to 50 K. This demonstrates that the spurion is magnetic in nature, even though it is not inelastic. We have not uniquely identified where the spurion comes from but it is likely to result from multiple scattering combining nuclear Bragg scattering with scattering by static moments.

Lastly, the error bars in all our figures represent plus or minus one standard deviation, using the standard rules for error propagation and with an error of $\pm\sqrt{\text{N}}$ assigned to a signal of N counts in the scattered signal as measured in the detector. The error bars obtained from fitting procedures were estimated using a Levenberg-Marquardt least squares fit to the scattering data (plus error bars) using a non-linear function. The datasets generated during and/or analysed during the current study are available from the corresponding author on reasonable request.

\bibliographystyle{apsrev4-1}

\bibliography{clusterdyn}

[floatfix]

\end{document}